\documentclass{emulateapj}
\usepackage{lineno}
\usepackage{psfig,enumerate}
\usepackage{color}
\usepackage{atbegshi}
\usepackage{url}
\usepackage[caption=false]{subfig}

\begin{document}


\title{White dwarf binaries Across the H-R Diagram}

\shorttitle{WD binaries in APOGEE}
\shortauthors{Anguiano et al.}

\author{Borja Anguiano\altaffilmark{1}, Steven R.\ Majewski\altaffilmark{1}, Keivan G.\ Stassun\altaffilmark{2}, Carles Badenes\altaffilmark{3}, Christine Mazzola Daher\altaffilmark{3}, Don Dixon\altaffilmark{2}, Carlos Allende Prieto\altaffilmark{4,5}, Donald P.\ Schneider\altaffilmark{6}, Adrian M.\ Price-Whelan\altaffilmark{7}, Rachael L.\ Beaton\altaffilmark{8}}


\altaffiltext{1}{Department of Astronomy, University of Virginia, Charlottesville, VA, 22904, USA}
\altaffiltext{2}{Department of Physics \& Astronomy, Vanderbilt University, 2301 Vanderbilt Place, Nashville, TN 37235, USA}

\altaffiltext{3}{Department of Physics and Astronomy and Pittsburgh Particle Physics, Astrophysics and Cosmology Center (PITT PACC),\\ University of Pittsburgh, 3941 O`Hara Street, Pittsburgh, PA 15260, USA}
\altaffiltext{4}{Instituto de Astrof\'isica de Canarias, E-38205 La Laguna, Tenerife, Spain}
\altaffiltext{5}{Universidad de La Laguna, Dpto. Astrof\'isica, E-38206 La Laguna, Tenerife, Spain}
\altaffiltext{6}{Department of Astronomy \& Astrophysics, Pennsylvania State University, University Park, PA 16802, USA}
\altaffiltext{7}{Center for Computational Astrophysics, Flatiron Institute, 162 5th Ave., New York, NY 10010, USA}
\altaffiltext{8}{The Observatories of the Carnegie Institution for Science, 813 Santa Barbara Street, Pasadena, CA 91101, USA}


\begin{abstract}

We created the APOGEE-GALEX-\emph{Gaia} catalog to study white dwarfs binaries. This database aims to create a minimally biased sample of WD binary systems identified from a combination of GALEX, {\it Gaia}, and APOGEE data to increase the number of WD binaries with orbital parameters and chemical compositions. We identify 3,414 sources as WD binary candidates, with nondegenerate companions of spectral types between F and M, including main sequence, main sequence binaries, subgiants, sub-subgiants, red giants, and red clump stars. Among our findings are (a) a total of 1,806 systems having inferred WD radii $R < 25$ R$_{\Earth}$, which constitute a more reliable group of WD binary candidates within the main sample; (b) a difference in the metallicity distribution function between WD binary candidates and the control sample of most luminous giants ($M_H < -3.0$); (c) the existence of a population of sub-subgiants with WD companions; (d) evidence for shorter periods in binaries that contain WDs compared to those that do not, as shown by the cumulative distributions of APOGEE radial velocity shifts; (e) evidence for systemic orbital evolution in a sample of 252 WD binaries with orbital periods, based on differences in the period distribution between systems with red clump, main sequence binary, and sub-subgiant companions and systems with main sequence or red giant companions; and (f) evidence for chemical enrichment during common envelope (CE) evolution, shown by lower metallicities in wide WD binary candidates ($P > 100$ days) compared to post-CE ($P < 100$ days) WD binary candidates.

\end{abstract}




\section{Introduction}

The evolution of multiple-star systems is a fundamental question in stellar astrophysics. Most stars in these systems are widely separated (i.e., on long period orbits), such that the pair does not interact strongly, and both stars evolve independently, as if single stars \citep{Willems2004}. However, around 25$\%$ of those are compact enough to exchange mass, changing the structures and subsequent evolution of both stars. For those systems with orbital periods less than $\sim$10 years --- ``close" binaries --- as the more massive component evolves from the MS to red giant branch (RGB), the stars undergo a stage of common envelope (CE) evolution, where mass can be transferred to the lower-mass MS star, and eventually leave behind a close, post-CE binary (PCEB) containing the core of the giant in the form of a white dwarf (WD) and its companion \citep[e.g.,][]{Webbink2008,Ivanova2013}.   

Physical understanding of CE evolution is extremely complicated from first principles, since there are too many length and time scales involved.  Moreover, the CE phase is very short, 400-4000 yrs \citep{Hjellming1991} and thus there are few known CE-phase candidates to provide observational constraints. For this reason, our primary strategy to understand the evolution of close binaries is to focus on the pre- and post-CE phases, and in particular on the obtention of systematic, unbiased and statistically robust surveys of systems in those evolutionary stages. Subsequent stages of compact binary evolution are no less interesting, as they lead to a variety of phenomena that play significant roles in numerous areas of astrophysics, from the creation of a diverse taxonomy of variable stars to the production of sources of gravitational waves and cosmological standard candles. PCEBs are the progenitors of many interesting astrophysical transients observed across the electromagnetic spectrum, including cataclysmic variables (CVs), novae, Type Ia SNe, and some core collapse SNe subtypes \citep[e.g.][and references therein]{Toloza2019}. 
As a result,
the properties of the binaries containing stellar remnants provide fundamental clues to 
understanding the varieties of subsequent evolution of PCEBs. However, the vast majority of these late evolutionary pathways start with a PCEB consisting of a WD and MS star, and, in all cases, constraining the detailed physics of these pathways relies on firm knowledge of the frequency and distribution of such systems by component masses, temperatures, separation, eccentricities,
and other properties. 
Because it is now clear that these fundamental statistics of stellar multiplicity 
are (a) strong functions of stellar properties like mass and chemical composition, and (b) not independent of each other \citep[e.g.,][]{badenes2018,Moe2019,mazzola2020}, the {\it only} way to fully characterize the rich phenomenology of PCEBs and their associated astrophysical transients is to assemble samples of systems with well-measured parameters that are large enough for robust multivariate statistical analysis. Furthermore, mergers of compact-star binaries are expected to be the most important sources for forthcoming gravitational-wave astronomy.

From an observational perspective, the Sloan Digital Sky Survey, SDSS \citep{York2000} was efficient at discovering new WDMS binaries \citep[e.g.,][]{Silvestri2006,Schreiber2008,Heller2009,Nebot2009} from optical spectra, but with biases towards systems containing hot WDs and secondaries of late spectral type. Moreover, in \cite{Corcoran2020} we found that some objects classified as WDMS in the 
SDSS sample are actually young stellar object contaminants. Nevertheless, to date, 
there are more than 3200 WDMS binaries found using SDSS spectra \citep[see][for a compilation of these objects]{RM2012}. Meanwhile, applying a similar approach to LAMOST \citep{Zhao2012} optical spectra \citep[e.g.,][]{Ren2018}, around new 1000 systems have been reported. However, some of these systems classified as
WDMS are actually WD-red giant pairs \citep{Corcoran2020}.  Nevertheless, the number of WD binaries with known, {\it non-MS} secondaries is very small, and this limits the ability to understand the panoply of possible fates of WDMS systems after the secondary star evolves off the MS. Apart from their biases and contamination, another challenge to using the existing databases is that, while they provide a large number of candidate WD binary systems, optical spectroscopic surveys typically offer only one radial velocity (RV) epoch, which does not enable characterization of the orbits.  Dedicated programs of spectroscopic follow-up have been motivated to address this problem, but the magnitude of the task has limited to a few hundred the number of systems with well defined orbital parameters \citep[e.g.,][]{Schreiber2008,Schreiber2010}, and, only $\sim$ 120 can be considered to be strong PCEB candidates \citep{Lagos2022}. 


In the vast majority of known WDMS systems the secondary companion is a low-mass M dwarf, since these are relatively easy to identify from optical colors and spectra. In recent years, there have been efforts to identify WD binaries with more massive FGK type companions \citep[e.g.,][]{Parsons2016,Hernandez2021}, by combining optical stellar spectroscopic surveys such as RAVE \citep{Steinmetz2020} and LAMOST with GALEX photometry \citep{Bianchi2017}.
Meanwhile, 
\cite{Parsons2016} obtained Hubble Space Telescope ultraviolet spectra for nine systems that confirmed that the photometrically observed UV excess in these systems is indeed caused, in all cases, by a hot compact companion. 
Clearly broadening wavelength sensitivity to the ultraviolet part of the spectral energy distribution (SED) greatly increases leverage in identifying and characterizing a broader variety of WD binary systems.  

Here we adopt a similar, but even more expansive approach to identifying WD binaries across the H-R diagram, with secondary companions with a broad range of spectral types and in virtually all phases of stellar evolution. Our goal is to make a new, large, and systematic search for compact binary star systems containing white dwarfs by harnessing information contained in
the spectroscopic catalog of the Sloan Digital Sky Survey's (SDSS IV, \citealt{Blanton2017}) near-infrared Apache Point Observatory Galactic Evolution Experiment (APOGEE) project \citep{2017AJ....154...94M}, cross-matched with data from the optical {\it Gaia} \citep{Lindegren2018} and ultraviolet GALEX \citep{Bianchi2017} space missions.
The combination of APOGEE's large $H$-band
spectroscopic data set with the UV photometry from GALEX allows us to identify main-sequence F, G, K, and early M
stars having significantly bluer GALEX ($FUV-NUV$) color than can be expected for single MS stars, which is key to our methodology (see \S\ref{identification}). In addition, APOGEE's deliberate focus on sampling evolved stars means that our survey contains WD binary systems with secondary stars amply representing all luminosity classes. Meanwhile, the {\it Gaia} database brings not only uniformly measured photometric measurements at optical wavelenghts, but critical astrometric measurements helpful 
for our analysis (e.g., see \S\ref{T_R_WD}). 
We add to the GALEX and {\it Gaia} photometry the infrared measurements from 2MASS and WISE, which further widens the wavelength range of our SED-fitting (see \S\ref{SED_sec}). 

By merging these various surveys, we have created the APOGEE-GALEX-\emph{Gaia} Catalog (AGGC) of candidate compact binaries containing WD stars. Numbering over 3,400 sources, the size of this catalog is comparable to that of previous WDMS catalogs, but includes secondary companions in the main sequence, subgiant branch, red giant branch, and red clump phases of evolution, as well as systems that occupy the MS-binary and subsubgiant regions of the H-R Diagram. Furthermore, it has been shown that WDs in close binaries can acquire non-degenerate envelopes that have radii up to 20 R$_{\earth}$ \citep{Sokoloski2006,Lewis2020,Washington2021}. 

The unique properties of the APOGEE catalog confer additional advantages to the AGGC. APOGEE's high resolution spectroscopy makes possible detailed, multi-element chemical abundance characterization of the binaries as well as the derivation of precise ($\sim100$ m s$^{-1}$) RVs. Moreover, APOGEE deliberately visited stars over multiple epochs spanning as long as a decade which makes it possible to infer orbital information (e.g., derived periods, eccentricities, masses, separations) on the binaries from the time series RV data.  The 252 WD binary systems in the AGGC for which full Keplerian orbits can be derived is already comparable to the previous number so characterized after monitoring campaigns, and the newly charaterized systems have been selected in an unbiased way (via standard APOGEE targeting) with secondaries that span almost the full range of stellar evolution.  The multi-epoch velocity information for the AGGC makes it a unique tool to explore the evolution of PCEB WD binary properties as the secondary star evolves from the MS all of the way to the red clump. 

The layout of the paper is as follows: \S\ref{catalog} provides an overview of the creation of the AGGC WD binary candidate catalog, while \S\ref{SED_sec} describes the use of the system SEDs to constrain empirically fundamental parameters, such as the effective temperature and the stellar radius of the
WD candidates. In \S\ref{sec:results} we explore some of the population distributions for various properties of the WD binary candidates in the AGGC, and the creation of the final sample. In \S\ref{CMD_sect} we demonstrate the usefulness of the newly created catalog of WD binary candidates to address a variety of astrophysical questions related to the evolution of close binaries. In particular, we explore variations in such properties as the WD temperature, the secondary star metallicity, the binary period, and fraction of close binaries as a function of the evolutionary stage of the secondary star. We also discuss the relevance and interpretation of our results. Finally, we draw some general conclusions from a preliminary analysis of the AGGC in \S\ref{conclusion}, and summarize our findings in \S\ref{summary}.

\section{Identification of White Dwarf Stellar Companions with APOGEE, GALEX, and Gaia}
\label{catalog}

The details of the construction of the AGGC and various checks of its veracity are described below.


\subsection{Initial Merging of the APOGEE, GALEX and \emph{Gaia} Catalogs}

The master database from which we will cull our APOGEE-GALEX-\emph{Gaia} Catalog (AGGC) starts with the APOGEE DR17 catalog \citep{sdss17}. The data were collected using the the SDSS telescope located at Apache Point Observatory \citep{Gunn2006}, the du Pont telescope \citep{BV1973} in Las Campanas, and the APOGEE spectrographs \citep{Wilson2019}. The APOGEE selection function for the main survey is simple, with generally only magnitude and color cuts  favoring  redder  stars \citep{Beaton2021,Santana2021}; as a  result, the APOGEE sample is mostly RGB stars ($\sim70\%$) and red MS stars ($\sim30\%$) \citep[e.g.,][]{Zasowski2017}. While these  criteria  prevent the discovery   of   individual white  dwarfs, the  large  number  of late  type MS  stars  in APOGEE are a rich parent sample  within which  to hunt for WD binaries.    

The entire APOGEE catalog has already been cross-matched with {\it Gaia} eDR3 \citep{Riello2020} and the 2MASS point source catalog \citep{Skrutskie2006} as part of SDSS DR17.  From the full APOGEE database \citep{Nidever2015}, we select only stars for which the APOGEE Stellar Parameters and Chemical Abundances Pipeline (ASPCAP) satisfactorily analyzed the combined spectrum of the source to obtain a basic set of stellar atmospheric parameters (e.g., $T_{eff}$, $[M/H]$, $\log{g}$) \citep{Garcia2016,Holtz2018}.\footnote{These are sources for which the APOGEE\_ASPCAPFLAG bitmask 23 is not set to ``BAD''.  The latter happens if any of the TEFF, LOGG, CHI2, COLORTE, ROTATION, SN error parameter bits are set to ``BAD'', or if any of the above parameters is near a grid edge in the ASPCAP synthetic spectral library used to derive atmospheric parameters (i.e., the flag GRIDEDGE\_BAD is set in any PARAMFLAG).  See the definitions of APOGEE flags as described at https://www.sdss.org/dr17/algorithms/bitmasks/.}

We cross-matched the APOGEE DR17 catalog against the GALEX database \citep{Bianchi2017}.  This results in 244,432 stars in common using a separation smaller than 2.5 arcsec on the sky. We then cross-match these stars against the WISE mission catalog \citep{Wright2010}, to procure mid-infrared photometry for our sources. After intersecting all of these catalogs, we create the AGGC as all cross-matched sources that have complete photometric information, defined as having valid entries for the two GALEX bands ($FUV$, $NUV$), the three {\it Gaia} bands ($G$, $G_{BP}$, and $G_{RP}$), the three 2MASS bands ($J$, $H$, $K_s$), and all four WISE bands ($W1$, $W2$, $W3$, $W4$). This results in a total of 242,896 objects. 

\subsection{Identification of White Dwarf Binary Candidates via UV Excess}
\label{identification}

The comparison of  ultraviolet (UV) photometric data, e.g, as represented by the (FUV-NUV) color from GALEX, to APOGEE-derived temperatures is a sensitive stellar probe for the presence of companions in the form of hot stellar remnants \citep[e.g.,][]{Bianchi2011}.

\begin{figure}[ht]
\includegraphics[width=1\hsize,angle=0]{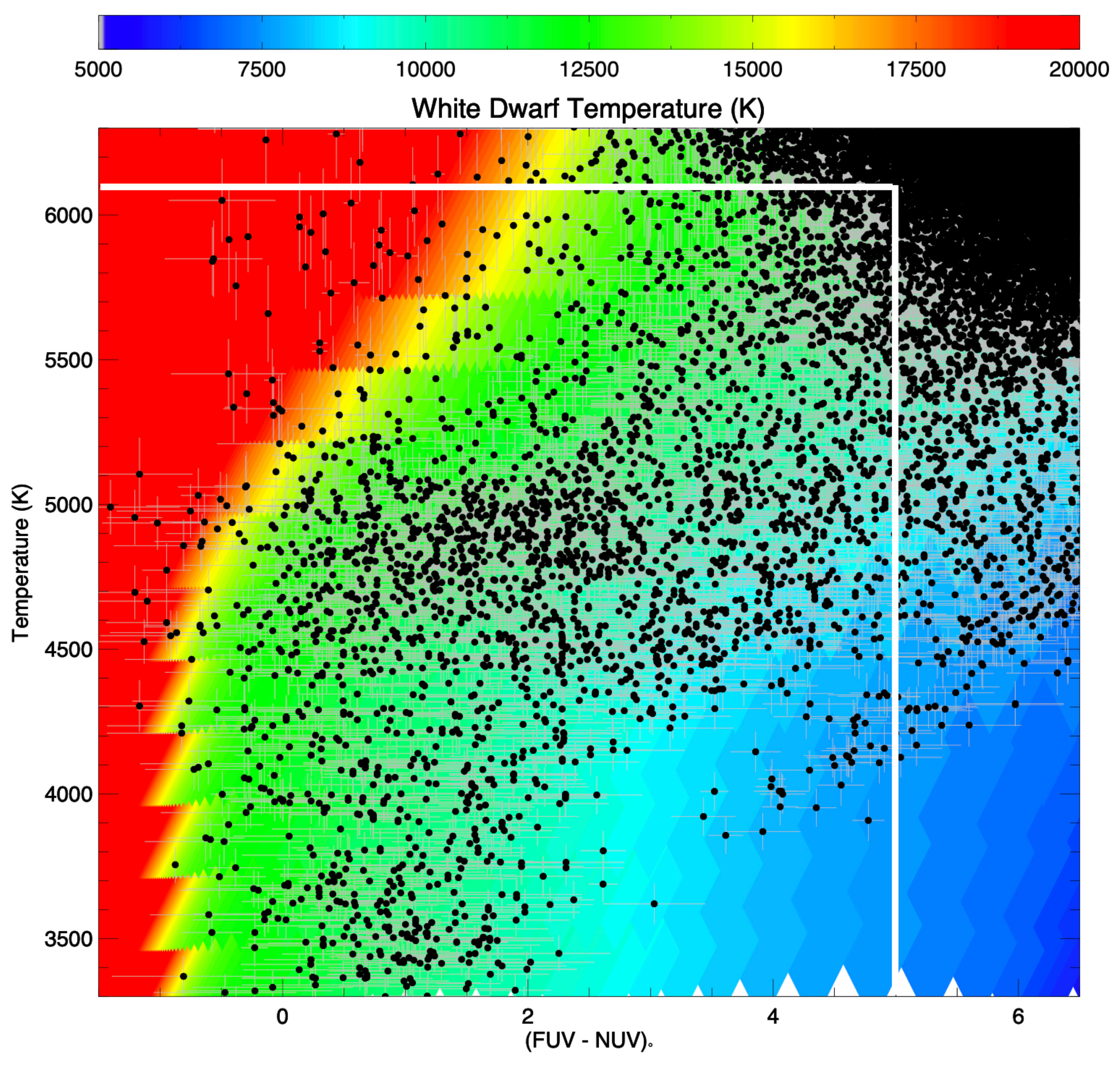}
\caption{APOGEE-derived effective temperature versus UV color for the selected stars having GALEX photometry. The figure is color-coded by the nominal WD effective temperatures inferred from ATLAS9 \citep{Castelli1997} and the WD \citep{Koester2010} grid of synthetic models for a range of temperatures, assuming main sequence stars as the red companion.
The white line shows the initial, simple selection criterion used for the AGGC, chosen to avoid the single star locus in the upper-right corner of the plot.}
\label{GALEX_APOGEE}
\end{figure}

Figure~\ref{GALEX_APOGEE}, which makes such a comparison, clearly shows stellar sources (black points) showing significantly blue UV colors, suggesting the presence of a hot source  --- a robust filter of candidate binaries with WDs \citep{Maxted2009,Morgan2012,Parsons2016}.
 
To demonstrate how effective is the combination of the GALEX UV color and the spectroscopically derived primary star effective temperature in the elucidation of WD companion properties, the background color scale of Figure~\ref{GALEX_APOGEE} shows what might be inferred for the WD effective temperature in the case of a hypothetical binary made from a solar metallicity, $\log{g} = 4.5$ MS star modeled with an ATLAS9 \citep{Castelli1997} atmosphere and a $\log{g} = 8$ WD modeled with energy distributions from the  \citet{Koester2010} grid of synthetic models.  To estimate the individual stellar fluxes for the stars we adopt a solar radius for the non-WD companion and an Earth radius for the WD.  This simple exercise immediately elucidates the WD effective temperature distribution we might expect within the AGGC: The largest number of the AGGC sources lie in modeled regions showing an inferred WD effective temperature range from 9,000 $<$ T$_{eff, WD}$ $<$ 15,000 K, while a few of the WD binary candidates show potential effective temperatures hotter than 20,000 K.\footnote{For cooler red star companions, the background color scheme in Figure~\ref{GALEX_APOGEE} is not significantly different in the case of a red giant companions of $\log{g} = 1.0$ and radius 10$R_{\odot}$, whereas when the red star temperature exceeds about 4500 K, the WD effective temperatures must be correspondingly hotter to result in the same ($FUV$$-$$NUV$) color.  Thus, the inferred WD effective temperatures from Figure~\ref{GALEX_APOGEE} represent something of a lower limit, for the case of a smaller red star companion.}





Based on the expected relative locations of single stars and WD binaries in the Figure~\ref{GALEX_APOGEE} parameter space \citep{Parsons2016,
Anguiano2020}, systems with
T$_{eff}$ $<$ 6000 K and (FUV - NUV)$_{\circ}$ $<$ 5 are the most compelling for our purposes.
Based on these criteria, we have identified 3,414 APOGEE sources that are WD binary candidates with F-M spectral type companions (see also Fig. 1 in \citealt{Anguiano2020}), and that ultimately constitute the final APOGEE-GALEX-\emph{Gaia} Catalog (AGGC) of WD binary candidates.  

The next step is to assess the robustness of this catalog and understand in more detail the types of objects it contains. In the next Section we describe how we can characterize the WD binary candidates via their spectral energy distributions.

\subsection{Understanding Sub-populations and Contaminants Among the White Dwarf Binary Candidates}

There are various types of stellar systems that can populate the Galex-APOGEE diagram. We can use previously characterized systems to understand these sub-populations as well as to vet the candidates and to identify sources of contamination. 

Indeed, one must be cautious that significant GALEX fluxes
are not fully reliable indicators of the presence of a WD, 
as there are other stellar sources that also emit significantly in the GALEX bands (e.g., various types of pulsating stars, chromospherically active stars, young stellar objects, subdwarfs) 
that could introduce false positive
WD binaries or complicate the interpretation of actual WD binaries.  In this Section, we explore these various issues that play a role in the reliability of the AGGC.

\subsubsection{WDMS Binaries in the SDSS}
\label{wdms_sdss}

To understand better our methodology
and what type of systems populate different parts of the Figure~\ref{GALEX_APOGEE} parameter space, we first look at the most up-to-date catalogs of well-established WDMS binaries
| the longest-lived, and therefore most common, binaries containing WDs | which are those identified in previous papers using SDSS optical spectra\footnote{sdss-wdms.org}, \citep[e.g.,][]{Schreiber2008,Nebot2009,RM2012}. The SDSS catalog contains a net total of 3,291 WDMS binaries, which, however, are affected by a mixture of selection effects.  Most significantly, because this catalog comprises sources wherein both the WD and MS stars are readily identifiable in the SDSS/SEGUE optical spectra, the SDSS sample is biased towards systems with the greatest temperature separation, namely, systems with relatively hot WDs and cool, M type dwarf companions.

\begin{figure*}[ht]
\includegraphics[width=1\hsize,angle=0]{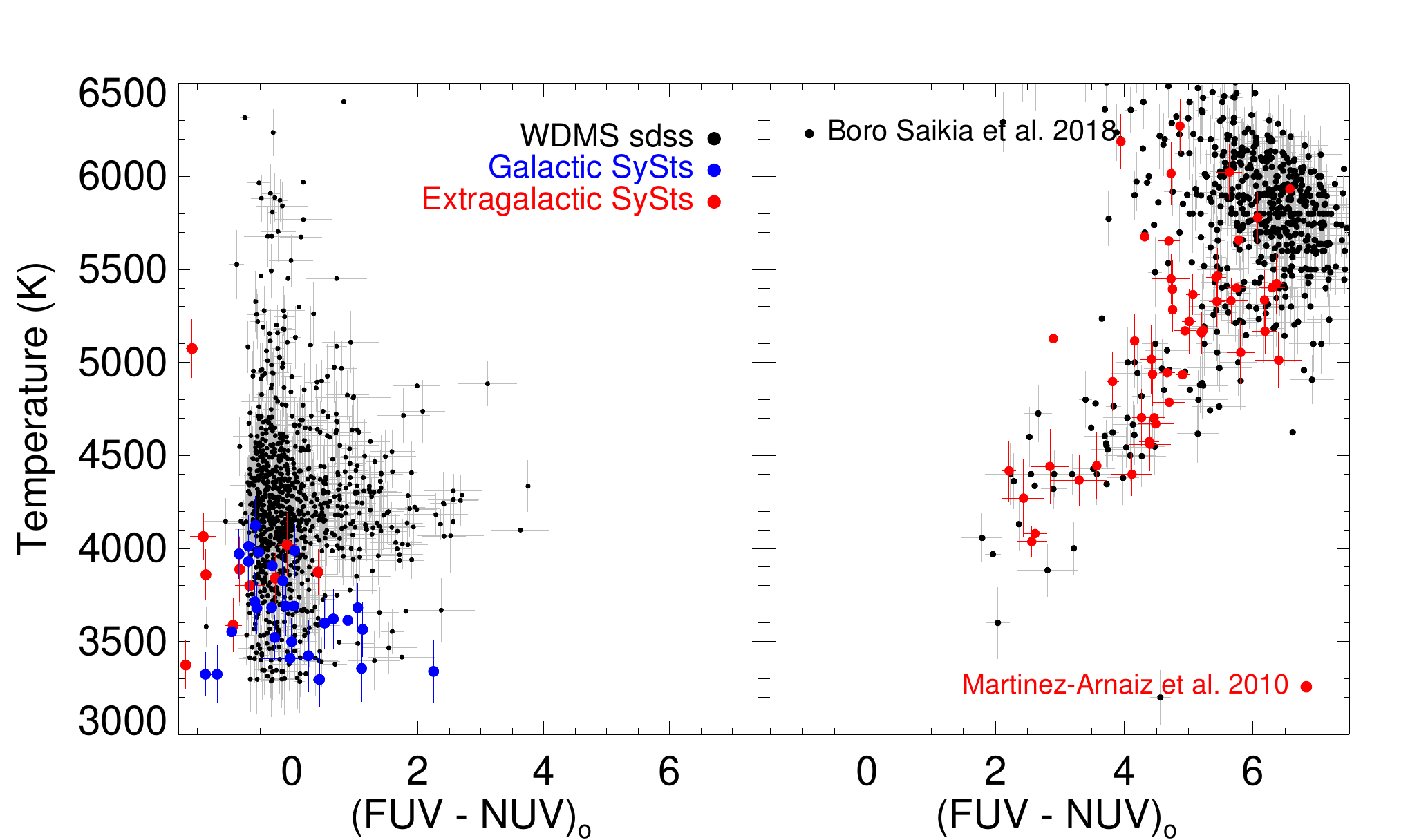}
\caption{The UV color vs temperature diagram for ({\it left}) WDMS binaries found in the SDSS/SEGUE survey, and the galactic and extragalactic symbiotic variables (SySts) reported in \cite{Merc2019}, and ({\it right}) chromospherically active stars explored by \citet{Boro2018} and \citet{MA2010}.
}
\label{GALEX_WDMS_ACTIVE}
\end{figure*}

To demonstrate how the more limited range of SDSS types of WDMS systems contrast with expectations from the AGGC, we cross-matched the SDSS WDMS catalog with GALEX and found 1,824 objects having $NUV$ and $FUV$ photometry available. We represent the UV color as a function of the effective temperature of the primary in the left panel in Figure~\ref{GALEX_WDMS_ACTIVE} for these objects. Nearly all of the WDMS pairs and the symbiotic binaries (SySts) show a UV color (FUV - NUV)$_{\circ}$ $<$ 2.0 and a MS star temperature lower than 4,800 K. Contrasting the left panel of Figure~\ref{GALEX_WDMS_ACTIVE} with Figure~\ref{GALEX_APOGEE} shows the greater senstivity to a broader range of primary and secondary stars expected in the AGGC.

\subsubsection{Stellar Chromospheric Activity}
\label{CA}

The parent catalog of WD binary candidates from the AGGC sample will contain potential contamination from various sources. For example, \cite{Amado1997} reported that single-lined spectroscopic RS Canum Venaticorum (RS CVn) systems, a variable type that consists of close binary stars having active chromospheres, show an ultraviolet excess. Hot chromospheres of active stars can result in them possessing quite blue UV colors \citep[e.g.,][]{Stelzer2013,Smith2018}. Furthermore, increased stellar activity in WDMS pairs where the MS is a M-dwarf have been reported in optical wavelengths by \cite{Morgan2012}, who proposed that such an increase in activity is a result of faster stellar rotation related to possible tidal effects, angular momentum exchange, or disk disruption (see also \citealt{Jones&West2016}).

\begin{figure}[ht]
\includegraphics[width=1.\hsize,angle=0,trim=30 10 20 20,clip]{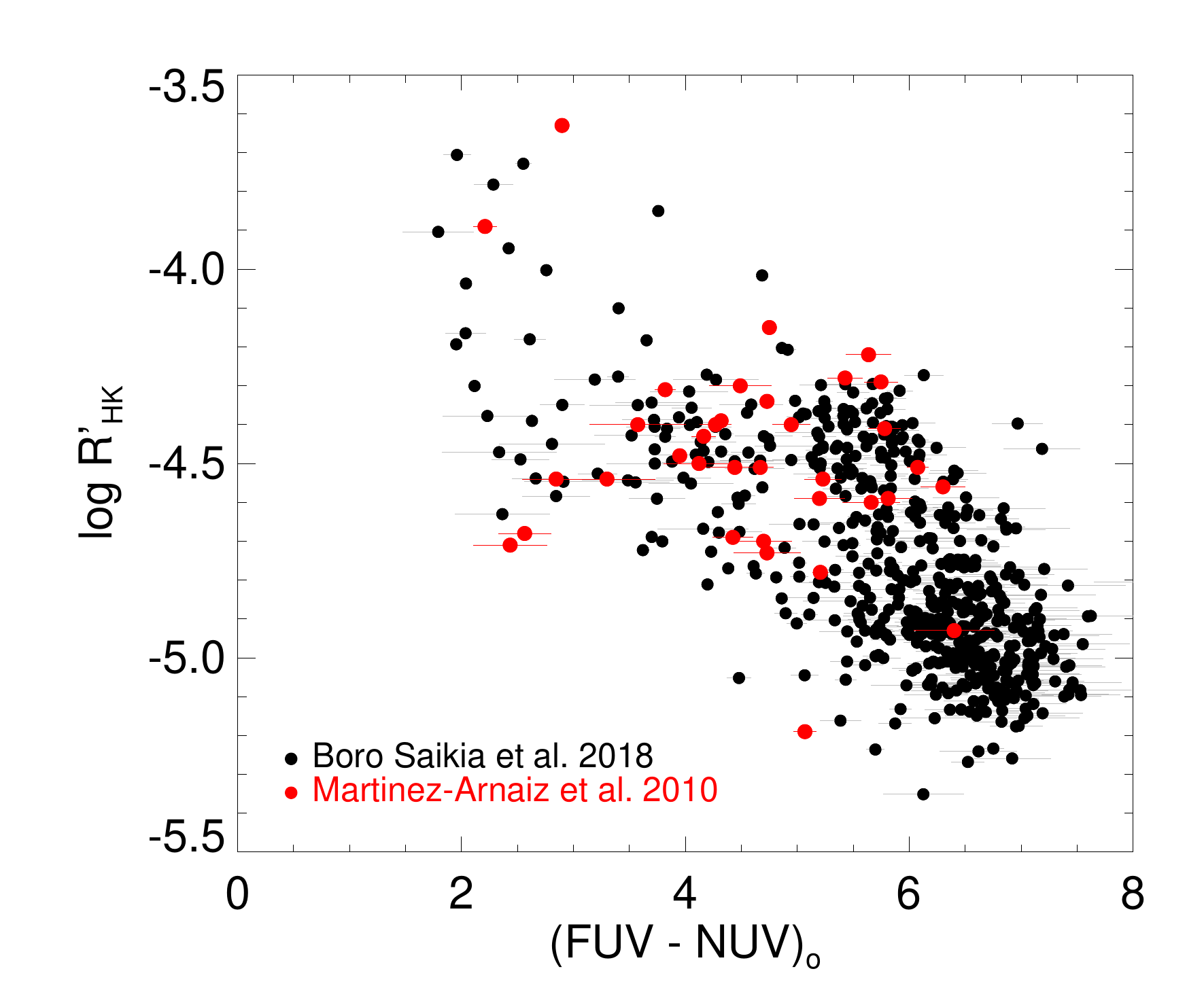}
\caption{The UV color versus $\log{R'_{\rm HK}}$ diagram for the \cite{Boro2018} (black dots) and \cite{MA2010} (red dots) catalogs. The most active stars tend to have UV colors $(FUV$$-$$NUV)_{\circ}<4.0.$}
\label{GALEX_logR}
\end{figure}

To investigate this potential source of contamination we include two catalogs of stars that, with spectroscopy at blue wavelengths, have detected Ca II H\&K line emission but with no obviously detected contribution of flux from a WD at optical wavelengths. \cite{Boro2018} presented a catalogue of chromospheric activity for 4,454 cool stars from a combination of archival HARPS spectra and several other surveys, including newly available data from Mount Wilson \citep{Baliunas1995}. In addition, \cite{MA2010} measured chromospheric activity, as given by different indicators throughout their optical spectra, and projected rotational velocities for 371 nearby cool single stars. For the latter study we select only stars classified as ``active".

The right panel in Figure~\ref{GALEX_WDMS_ACTIVE} shows the effective temperature versus UV color for the two chromospheric activity catalogs mentioned above. We observe a ``cool dwarf branch" that shows ($FUV$$-$$NUV$)$_{\circ}$ $<$ 5.0 and $T_{\rm eff}$ $<$ 5000 K. To gain insight into the origin of these stars, in Figure~\ref{GALEX_logR} we show the UV color with respect to the ratio of chromospheric Ca II H\&K flux to
bolometric flux, $\log{R'_{\rm HK}}$, for the \cite{Boro2018} and \cite{MA2010} catalogs (black and red dots, respectively, in each of Figs.~\ref{GALEX_WDMS_ACTIVE} and \ref{GALEX_logR}).  We observe that nearly all the stars with a color ($FUV$$-$$NUV$)$_{\circ}$ $<$ 5.0 are catalogued as chrosmopherically active stars. We also observe that activity is the primary driver of the UV color, where main-sequence stars cooler than $\sim$5000 K have at least some amount of chromospheric activity, while most of the non-active stars show a ($FUV$$-$$NUV$)$_{\circ}$ $>$ 6.0. This assessment suggests that the AGGC may have contamination from chromospherically active stars.
On the other hand, it may be significant that we do not find active stars in these two independent catalogs with ($FUV$$-$$NUV$)$_{\circ}$ $<$ 2.0, whereas the bulk of the AGGC catalog either has colors bluer than this or blueward of the Figure~\ref{GALEX_logR}) chromospherically active sequence.
Moreover, a star can be chromospherically active {\it and} have a WD companion.
To help further navigate this complicated range of possibilities, we will bring to bear the estimates of stellar radii from the UV GALEX and IR-bands together with parallaxes from \emph{Gaia} eDR3 (\S\ref{SED_sec}).


\subsubsection{Variable Stars}
\label{SIMBAD}

Pulsating stars like Cepheids and RR Lyrae can have blue GALEX ($NUV$$-$$FUV$) colors \citep[e.g.,][]{Welsh2005,KB2014}, and they can be a potential contamination in our candidate WD binaries sample.  We cross-matched our AGGC sample with a catalog 
of multi-band, time-series photometric characterisation of Cepheids and RR-Lyrae using \emph{Gaia} DR2  \citep{Clementini2019}; this catalog contains 150,359 such variables (9,575 classified as Cepheids and 140,784 as RR Lyrae stars) distributed across the sky. However, the catalog reaches to the \emph{Gaia} faint-magnitude limit of $G$ $\sim$ 20.7, well beyond APOGEE's nominal limits. As a result, we found only five RR Lyrae and four Cepheids in our AGGC sample identified in the \emph{Gaia} catalog.

\begin{figure*}[ht]
\includegraphics[width=1\hsize,angle=0,trim=0 20 20 30,clip]{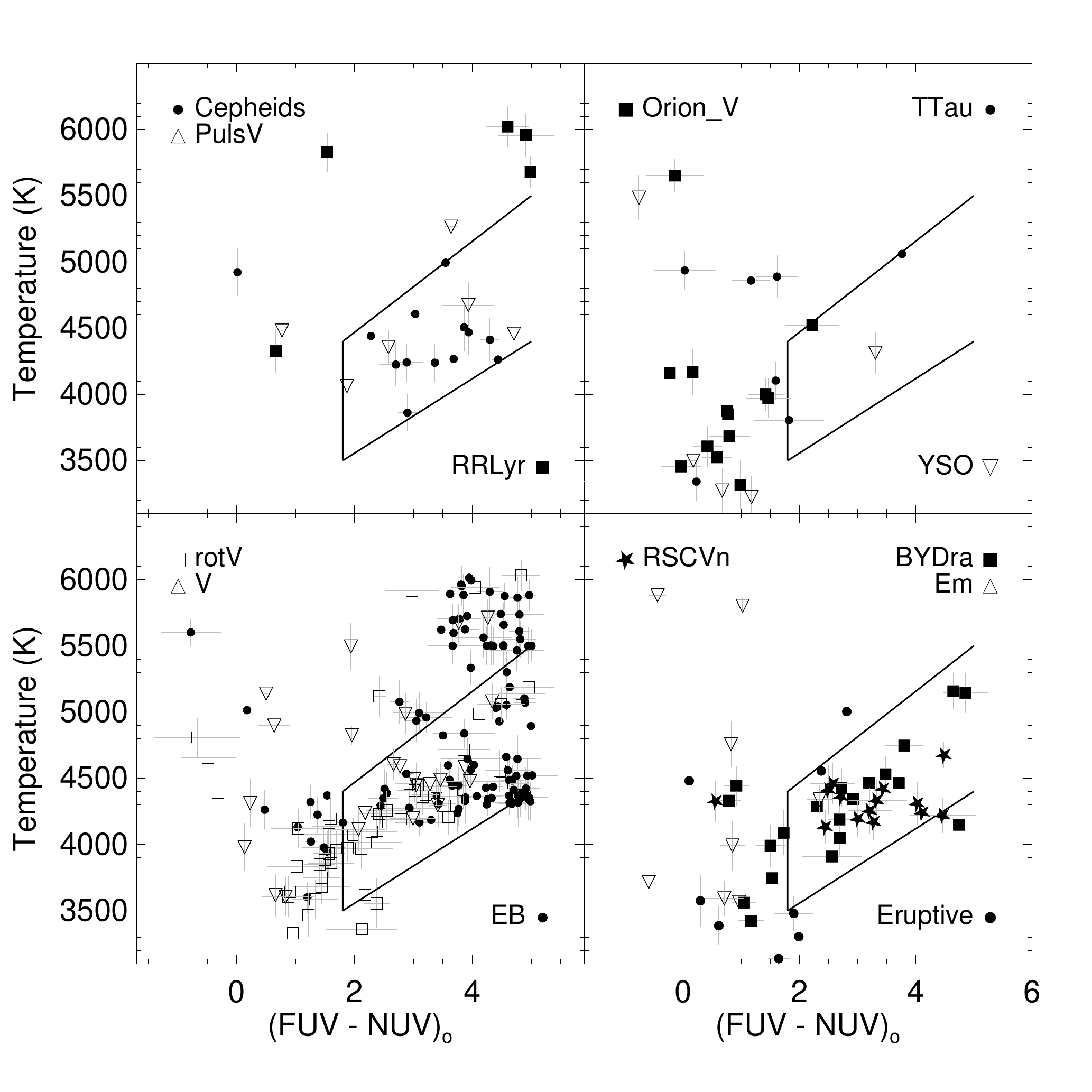}
\caption{The UV color versus APOGEE-derived temperature diagram for different types of stars as classified by \citet{Clementini2019} or as listed in the SIMBAD database. The solid lines inside the figures represent the ``chromospherically active area''. See text for details.}
\label{SIMBAD}
\end{figure*}

We also cross-matched the candidate WD binaries in the AGGC sample against the SIMBAD database \citep{Wenger2000}. The vast majority of our system candidates are classified as ``Star" or ``High Proper-motion Star", which demonstrates our overall ignorance about these objects (but also that the majority have not been found or studied previously).  However there are six candidate WD binaries classified as ``Symbiotic", and four as ``Dwarf Nova". Together with the Cepheids and RR Lyrae stars mentioned above, we also find six stars that were previously labeled as ``Pulsating Variable Star (PulsV)". In Figure~\ref{SIMBAD} we show the UV color versus $T_{\rm eff}$ diagram with four different panels to explore how specific type of stars currently in the SIMBAD archive occupy this parameter space.
The outlined region in the four panels shows the ``cool dwarf branch" for chromospherically active stars discussed above (\S\ref{CA}).  For example, the upper-left panel illustrates the distributions of Cepheids (black dots), RR Lyrae (black squares) and PulsV stars (triangles) in our candidates WD binary sample. 
Interestingly, most of the Cepheids and PulsV lie inside the ``active branch", whereas the RR Lyrae lie outside this region and are predominantly hotter than 5500 K. We also find 7 of our systems to be classified as ``T Tau-type star" or ``Candidate T Tau-type star".
In Figure~\ref{SIMBAD} (top-right panel) we show these TTau (black dots), along with ``Young Stellar Objects (YSO)" (triangles) and ``Variable Star of Orion Type (Orion-V)" (squares). All of these objects are very young stars in the pre-main sequence stage \citep[e.g.,][]{Mathieu1991,Kounkel2018}. Despite their UV color (predominantly $FUV - NUV < 1.5$), it is very unlikely that any of these systems have white dwarf companions \citep{Corcoran2020}. 

The bottom-left panel of Figure~\ref{SIMBAD} shows AGGC objects classified in SIMBAD as ``Eclipsing binaries (EB)" (black dots; 77 objects), ``Rotationally variable Stars (RotV; 33 objects)" (squares) and ``Variable Stars" (triangles; 22 objects). Most of the stars labeled as EB and RotV are within the ``chromospherically active area". Many contact binaries show activity signals, where the binary consists of two low-mass main sequence stars \citep[mostly F, G, or K spectral type; e.g.,][and references therein]{Mitnyan2020}. Moreover, we observe that some EB and RotV stars could have very blue UV colors, where $(FUV-NUV)_{\circ}\sim1.0$. Furthermore, 14 objects in our sample are listed in SIMBAD as ``RS Canum Venaticorum variable (RS CVn)" (star symbols in the lower right-hand panel of Figure~\ref{SIMBAD}). RS CVn are class of detached binary typically composed of a chromospherically active G or K stars \citep[e.g.,][]{Biazzo2006}. However, objects classified as RS CVn,  with strong Ca II H \& K lines in emission, where the hotter component is a WD are also reported \citep[e.g.,][]{Vaccaro2015}. All of the RS CVn but one are inside of the ``chromospherically active area" in the color-temperature diagram. We also have 20 objects listed as ``Variable of BY Dra type (BYDra)"; these are a class of object where light variability is caused by axial rotation of a star with a variable degree of non-uniformity of surface brightness due to, e.g., starspots and/or chromospheric activity \citep[e.g.,][]{Alekseev2000}. Apparently a close companion is a sufficient, but not a necessary, condition for the occurrence of the BY Dra phenomenon \citep{Bopp_Fekel1977,Eker2008}. Right-hand panel of Figure~\ref{SIMBAD}) also listed objects classified as Emission-line Star (EM, 8 objects). These objects are potential Novae or cataclysmic variables stars (CVs), where we see radiation coming from the accretion disks \citep[e.g.,][]{Idan2010}. Finally, we also show objects labeled as Eruptive in SIMBAD as black dots symbols. Some of these variable stars show flares and the changes in luminosity coincide with shell events or mass outflow in the form of stellar wind \citep[e.g.,][]{Tapia2015}. 


\section{Physical Properties of White Dwarf Companions via Spectral Energy Distributions}
\label{SED_sec}

Broadband photometric spectral energy distributions (SEDs) spanning from the ultraviolet to the infrared provide a means both to confirm the presence of a WD in our binary candidates and to constrain empirically the fundamental parameters of the two stars in each system, most importantly the WD effective temperature, each stellar radius. In this regard, the ultraviolet fluxes are critical for constraining the properties of the WD, since it is at these wavelengths that the WD typically dominates the SED. 


\subsection{Overall Approach}
\label{T_R_WD}

The empirical spectral energy distribution (SED) for the AGGC sample is an  aid not only in identifying 
WD binaries, but also in deriving system parameters like the WD effective temperature and radius. To create the SEDs, we use the photometric bands from GALEX \citep{Bianchi2017}, Gaia eDR3 \citep{Riello2020}, 2MASS \citep{Skrutskie2006} and WISE \citep{Wright2010}. Because APOGEE provides reliable information on the red star contributing to the SED, in principle we can use the residual flux to determine the properties (temperature and radius) of the WD.
To this end, following the methods laid out by \cite{ST2016}, we account for the red portion of the SED (after correcting for extinction by dust) 
with a Kurucz stellar atmosphere model \citep[e.g.,][]{Castelli1997} where we adopt the stellar atmospheric parameters reported by APOGEE \citep[e.g.,][]{Holtz2018}. 
If it is assumed that the residual SED flux is attributable to the WD, the properties of that star can be determined by a comparison of that measured residual energy distribution with the hot star spectrum predicted from an appropriately matching model atmosphere. 
In this case, to interpret the contribution of the WD to the net SED, we use the model WD spectra of \cite{Koester2010}.

Once the SED is adequately described, it is possible to ascertain the radii of the constitutent stars.  For those parts of the SED dominated by one or the other star in the binary, the radius, $R$, of the corresponding star in units of solar radii is given by conservation of flux, e.g., 
\begin{equation}
R = 4.43 \times 10^{7} r (F_{\lambda}/F_{\lambda, \mathrm{surface}})^{1/2}
\label{radius}
\end{equation}
\noindent where $r$ is the star's distance in parsecs, $F_{\lambda}$ is the apparent monochromatic flux, and $F_{\lambda, \mathrm{surface}}$ is the absolute flux at the surface of the dominantly contributing star  \citep[see][for an application of this approach in WDs]{Shipman1979}. In this case, the $F_{\lambda}$ are actually obtained using VOSA\footnote{http://svo2.cab.inta-csic.es/theory/vosa/} \citep{Bayo2008} for the specific photometric bands mentioned above, while to compute the absolute flux at the surface of the star, $F_{\lambda,surf}$, we use (as mentioned above) either the model WD atmosphere spectra of \cite{Koester2010} or the Kurucz stellar atmosphere model \citep[e.g.,][]{Castelli1997}, as appropriate.
In practice, to measure the WD radii we use the fluxes from the two GALEX bands, while to measure the radius of the secondary we use the fluxes from the 2MASS bands along with the W1 and W2 bands from WISE. 



Note that the SED fitting procedure implicitly assumes that both the cool component and the hot component are single objects. This assumption naturally breaks down for some types of systems, such as triple systems where the cool component is in fact itself an unresolved binary. The most extreme case is that for which the cool component is an equal-mass binary, such that the observed flux is a full factor of 2 larger than for a single star. In that case, the cool component will appear to have an inferred radius that is over-estimated by a factor of $\sqrt{2}$. As we discuss in Section~\ref{sec:results}, only 192 systems are affected. Most importantly this does not impact the derived parameters for the WD component, unless the system is a triple with two WDs, which is extremely unlikely.

\begin{figure}[ht]
\includegraphics[width=\hsize,angle=0,trim=30 10 20 20,clip]{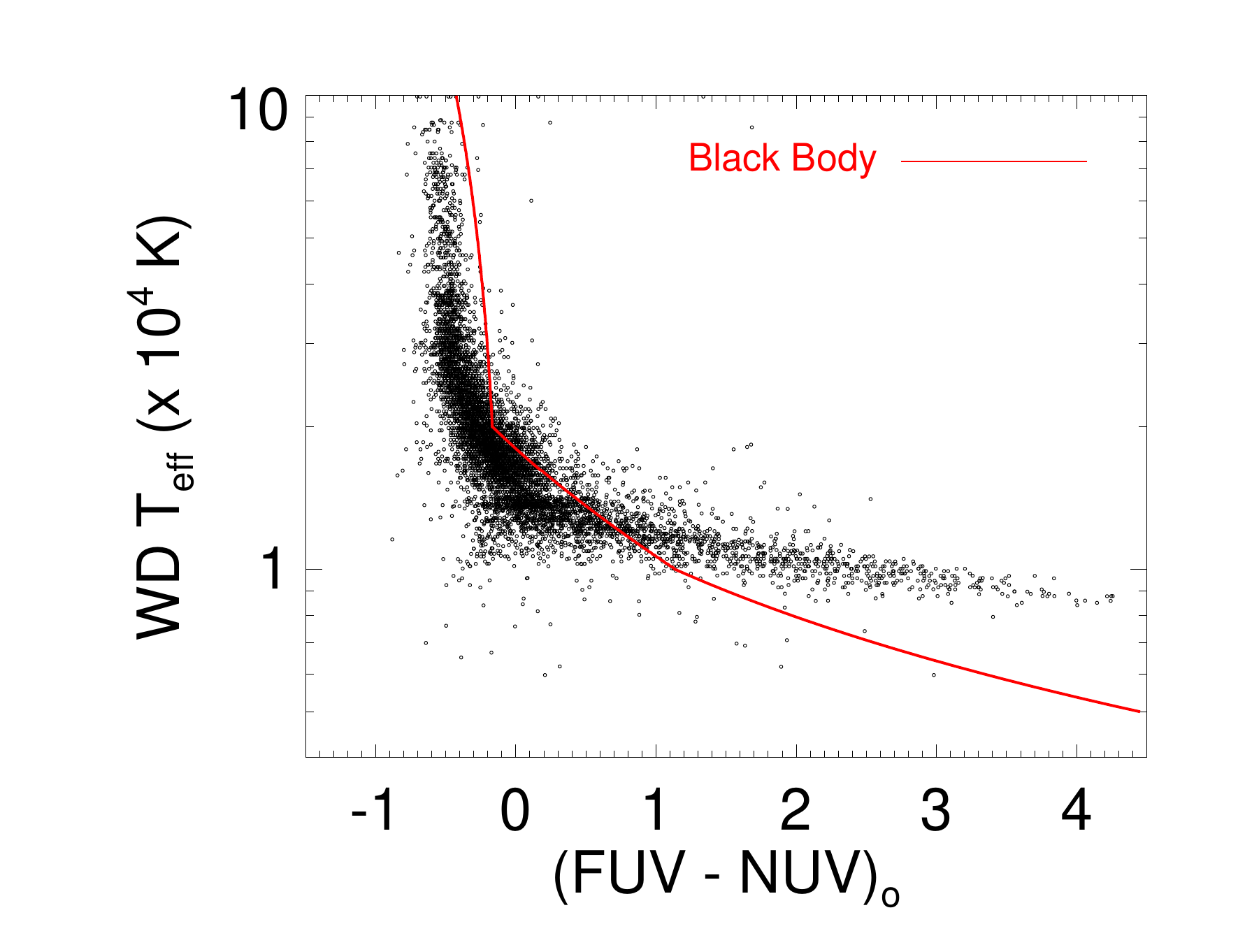}
\caption{The dereddened UV color vs temperature distribution for confirmed white dwarfs in the SDSS.  The red line shows expectations for a blackbody. 
{The blackbody was generated by the synthetic models in the Virtual Observatory facility (VOSA), the steps in temperature jump from 50 K to 1000 K for temperature hotter than 20000 K; the log scale amplifies this effect.}
}
\label{WD_temp}
\end{figure}

The effective temperature of the possible hot companion in our system candidates is another parameter we can extract from the SED analysis. This temperature is the primary driver of the GALEX UV color. To understand the relation between the ($NUV$$-$$FUV$) color and the temperature of the WD, we use the SDSS DR12 WD catalog with spectroscopic temperatures used in \cite{Anguiano2017} to create the distribution in Figure~\ref{WD_temp}.
In this case, the WD effective temperatures were derived \citep{Kepler2016}
by fitting the Balmer lines sampled by the SDSS spectra with the one-dimensional model atmosphere spectra of \cite{Koester2010}.  However, we note a 
discrepancy between the Kepler et al.-derived color-temperature relation for the SDSS WDs and expectations from a blackbody (red line in Fig.~\ref{WD_temp}), with
those differences increasing in the cool regime. We use the synthetic photometry and temperature listed in VOSA \citep{Bayo2008} for the blackbody in the UV-bands. This discrepancy could be related to the effect on temperature of convective atmospheres, 
which is important for cool white dwarfs \citep[e.g.,][]{T2013}, and demonstrates that invoking a blackbody to represent the WD contribution to the SED may lead to systemic offsets in derived temperatures. 
As may also be seen in Figure~\ref{WD_temp},
the UV GALEX color saturates to a nearly constant value 
for WDs with $T_{\rm eff}$ $>$ 30,000 K.  While this may caution one from trusting SED-fitting at such hot temperatures, our procedure here, which makes use of actual WD models rather than simple blackbodies, does at least produce results for the hottest WDs (see Fig.~\ref{UV_color_TEFF} below) that are more in line with the \cite{Kepler2016} methodology shown in Figure~\ref{WD_temp}.

\begin{figure}[ht]
\includegraphics[width=\hsize,angle=0,trim=20 10 20 20,clip]{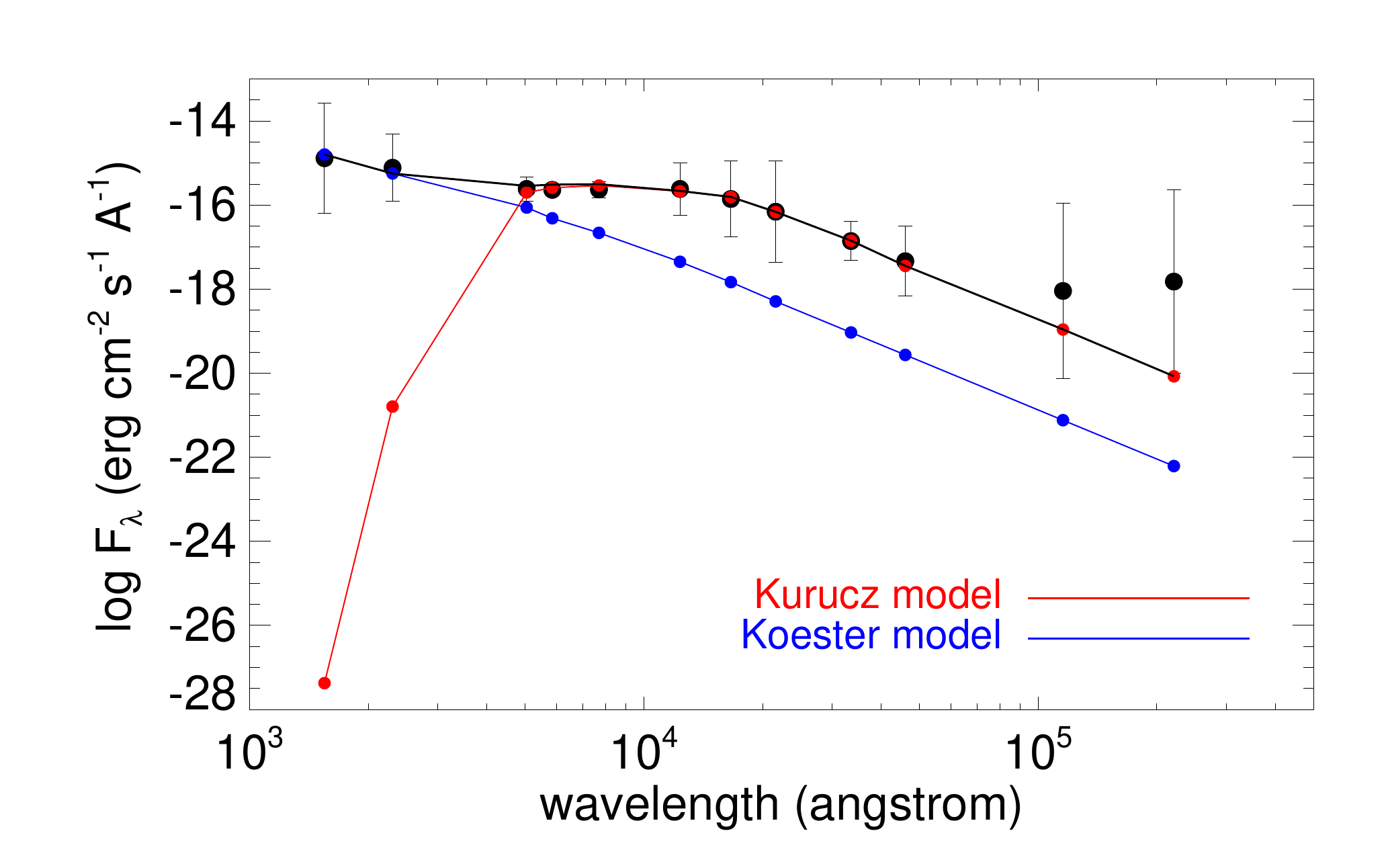}
\includegraphics[width=\hsize,angle=0]{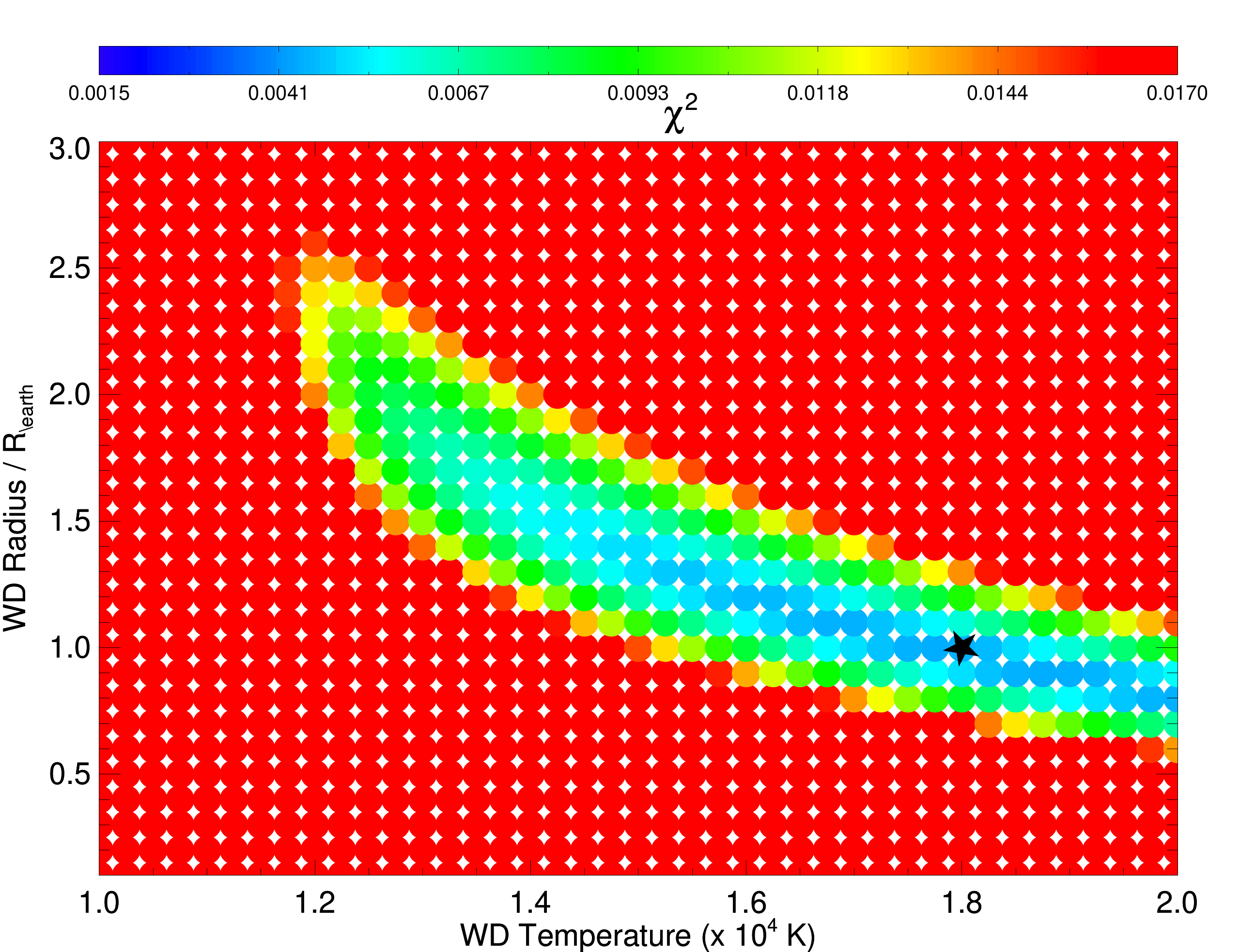}
\caption{\emph{Top panel:} The best-fitting spectral energy distribution ({\it black line}) for a WDMS binary using GALEX, {\it Gaia}, 2MASS and WISE broadband photometry ({\it black dots}). The red line is the best Kurucz model \citep{Castelli1997} fit for the secondary, while the blue dots/line represents the best-fitting \citet{Koester2010} model for the WD.
\emph{Lower panel:} WD Temperature - Radius plane color-code by the $\chi^2$ value. The blue color indicates the lowest $\chi^2$ value   }
\label{SED_chi}
\end{figure}

\begin{figure*}[ht]
\includegraphics[width=1.\hsize,angle=0]{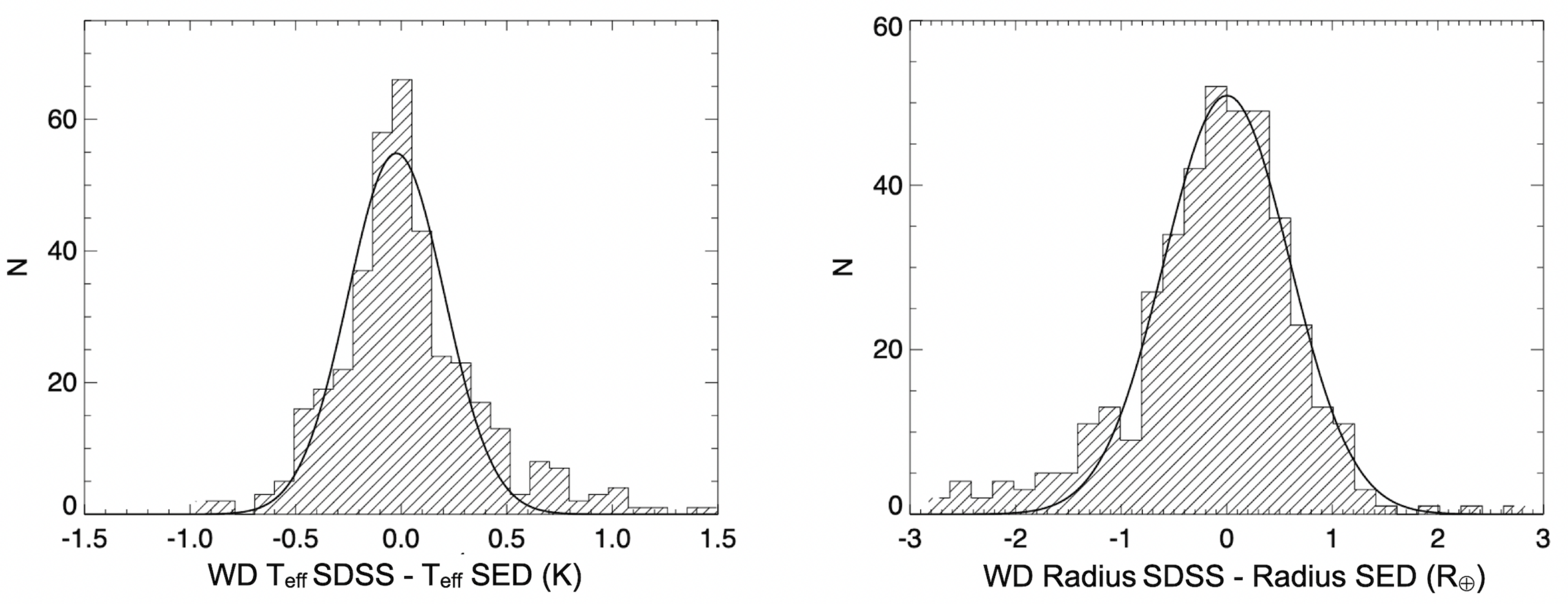}
\caption{Histogram of the discrepancies between the temperature and the radius derived in the SDSS spectra and those derived in this work using an SED fitting procedure. A fit with a simple Gaussian (black line) reveals small offsets in the results coming from the two methods and a dispersion of $\sim$2200~K for the temperature and $\sim$0.6~R$_{\Earth}$ for the WD radius. 
}
\label{SED_sdss}
\end{figure*}

In the end, we derived the WD radii and effective temperatures for each candidate WD binary by simultaneously fitting to the observed SED the combination of a single, APOGEE-motivated red star model and a variety of blue star model contributions, where temperature and radius are the free parameters.
The best fit is found through $\chi^2$ minimization, where the latter is given by 
\begin{equation}
    \chi^2 = \sum_{i=1}^{n} \frac{(O_{i} - E_{i})^2}{E_{i}}
\end{equation}
\noindent where $E_{i}$ is the sum of flux from the models for the two stars, $O_{i}$ is the SED, and where we estimate the agreement between the observed and the expected distributions for all ten of the individual photometric bins, $i$, utilized (see below). 


While for the secondary star we adopt the stellar parameters ($T_{\rm eff}$, [M/H], $\log{g}$) given by APOGEE DR17, in practice the SEDs are primarily sensitive to $T_{\rm eff}$, with little or no influence by (i.e., and therefore little sensitivity  to) the overall stellar metallicity or surface gravity. In the fitting of the secondary star SED we use all the photometric bands we excluded use of the WISE W3 and W4 bands, which often show a clear deviation from the models for the nominal secondary star temperature (see the top panel in Fig.~\ref{SED_chi} for a typical example). Infrared (IR) excesses observed in these bands may suggest the presence of warm dust around the star \citep[e.g.,][and references therein]{DaCosta2017}. For the SED extinction correction, we deredden the observed flux using the reddening curve from \cite{Fitzpatrick1999} parameterization, which is valid from the IR to the far-UV. The $E(B-V)$ value is given for each source in the GALEX catalog \citep{Bianchi2017}, based on the extinction maps of \cite{Schlegel1998}. A typical ``average" extinction law for the diffuse interstellar medium, where $R(V) = 3.1$, is assumed. 


\begin{figure*}[ht]
\includegraphics[width=1.\hsize,angle=0]{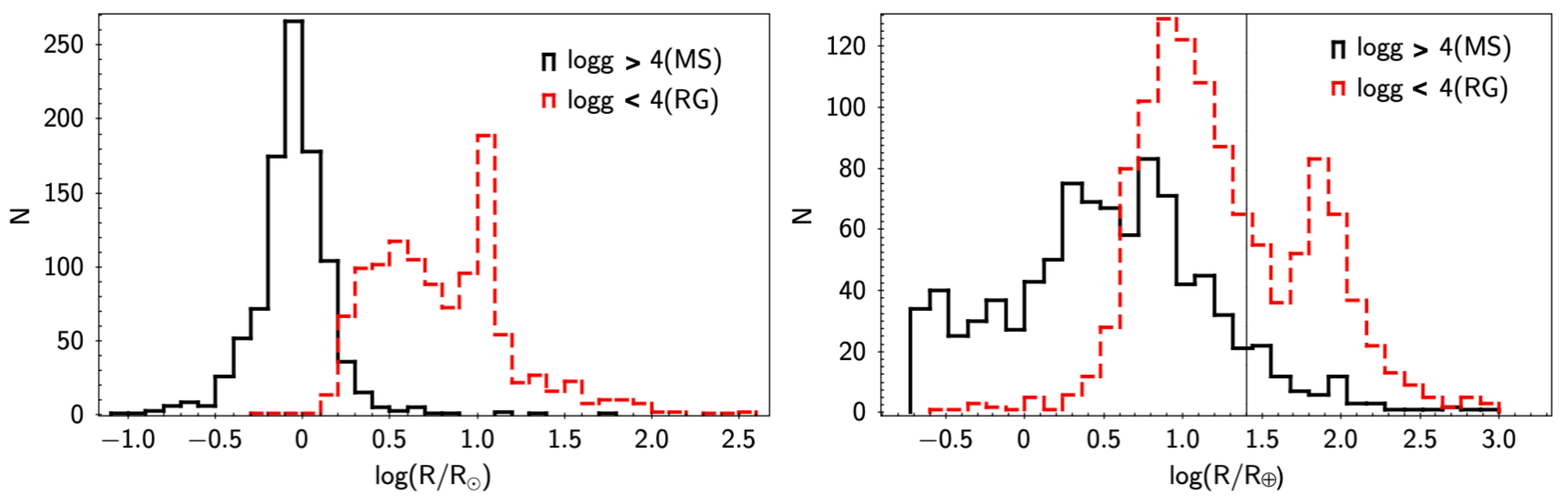}
\caption{\emph{Left panel:} Stellar radius logarithmic distribution for the MS and RG star secondaries in solar radii. Most of the MS stars have radius close to 1 R$_{\odot}$ (black line). We also have sub-giants and giant star secondaries with a large range in radius (red dashed line). We use the APOGEE surface gravity to select MS and RG objects, as indicated in the legend. 
The distribution shows a group of giants with $R$ $\sim$ 10 R$_{\odot}$, associated with the RC. \emph{Right panel:} Stellar radius distribution for the WD candidates, where the units are in Earth radii for MS (black line) and RG (red dashed line). The vertical black line represents the value $R$ = 25 R$_{\earth}$. The number of WDs with $R < 3$ R$_{\earth}$ is much larger for MS than for RG. In the RG distribution most of the WDs have a $R \sim 9$ R$ _{\earth}$. }
\label{radius_distribution}
\end{figure*}

\begin{figure}[ht]
\includegraphics[width=1.\hsize,angle=0]{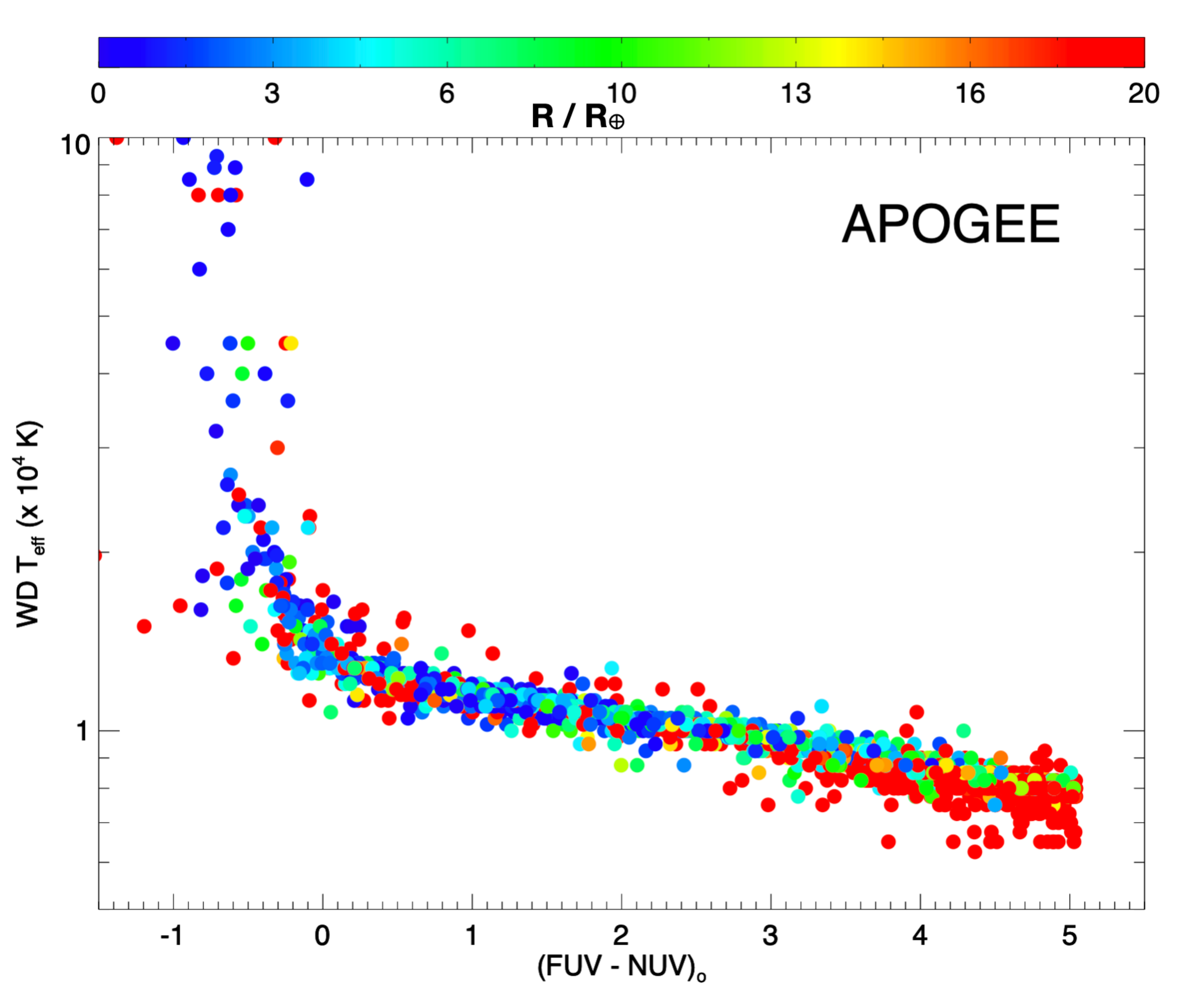}
\caption{Comparison between the WD temperature derived using SED-fitting and the ($FUV - NUV$) GALEX color. The figure is also color-coded with the WD radii. Stars with $R_{WD}$ $>$ 20 R$_{\Earth}$ tend to have $T_{\rm eff}$ $<$ 10$^{4}$ K.}
\label{UV_color_TEFF}
\end{figure}

\begin{figure}[ht]
\includegraphics[width=\hsize,angle=0]{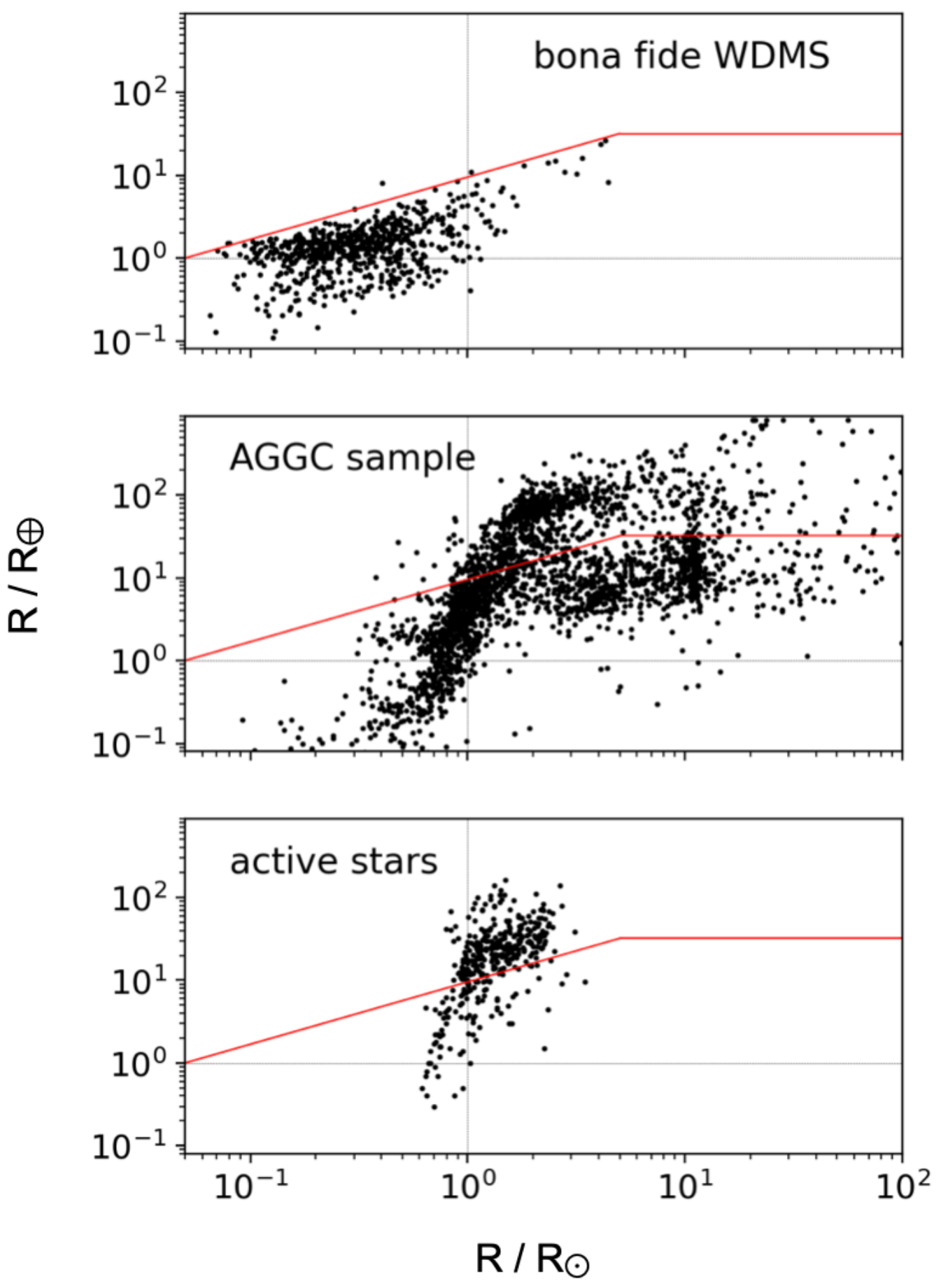}
\caption{APOGEE-derived radii for the non-WD  stars  versus the nominal  WD  radii  inferred  from  SED-fitting  (\S2.4) for: ({\it top}) SDSS optical WDMS candidates (sdss-wdms.org); ({\it middle}) our  parent  sample of WD binary candidates based on Fig.~\ref{GALEX_APOGEE}; and ({\it lower}) chromospherically active stars from \citet{Boro2018}. In each panel, the red line represents a criterion, guided by and separating the loci of most single stars, most chromospherically active stars, and almost all {\it bona fide} WD stars from SDSS; the best AGGC stars will lie below this line.}
\label{WD_radii_MS}
\end{figure}

\begin{figure*}[ht]
\includegraphics[width=.48\hsize,angle=0]{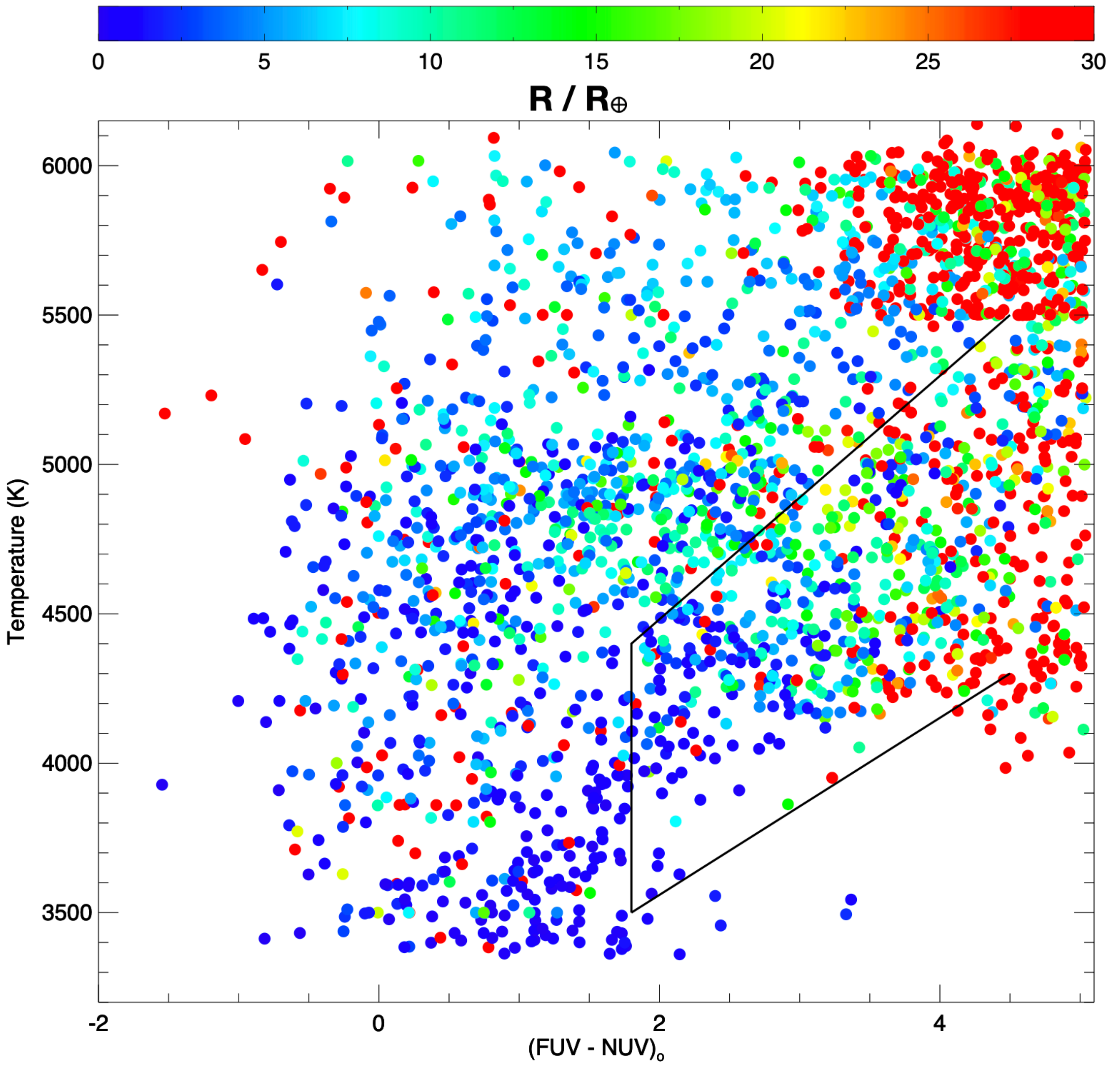}
\includegraphics[width=.48\hsize,angle=0]{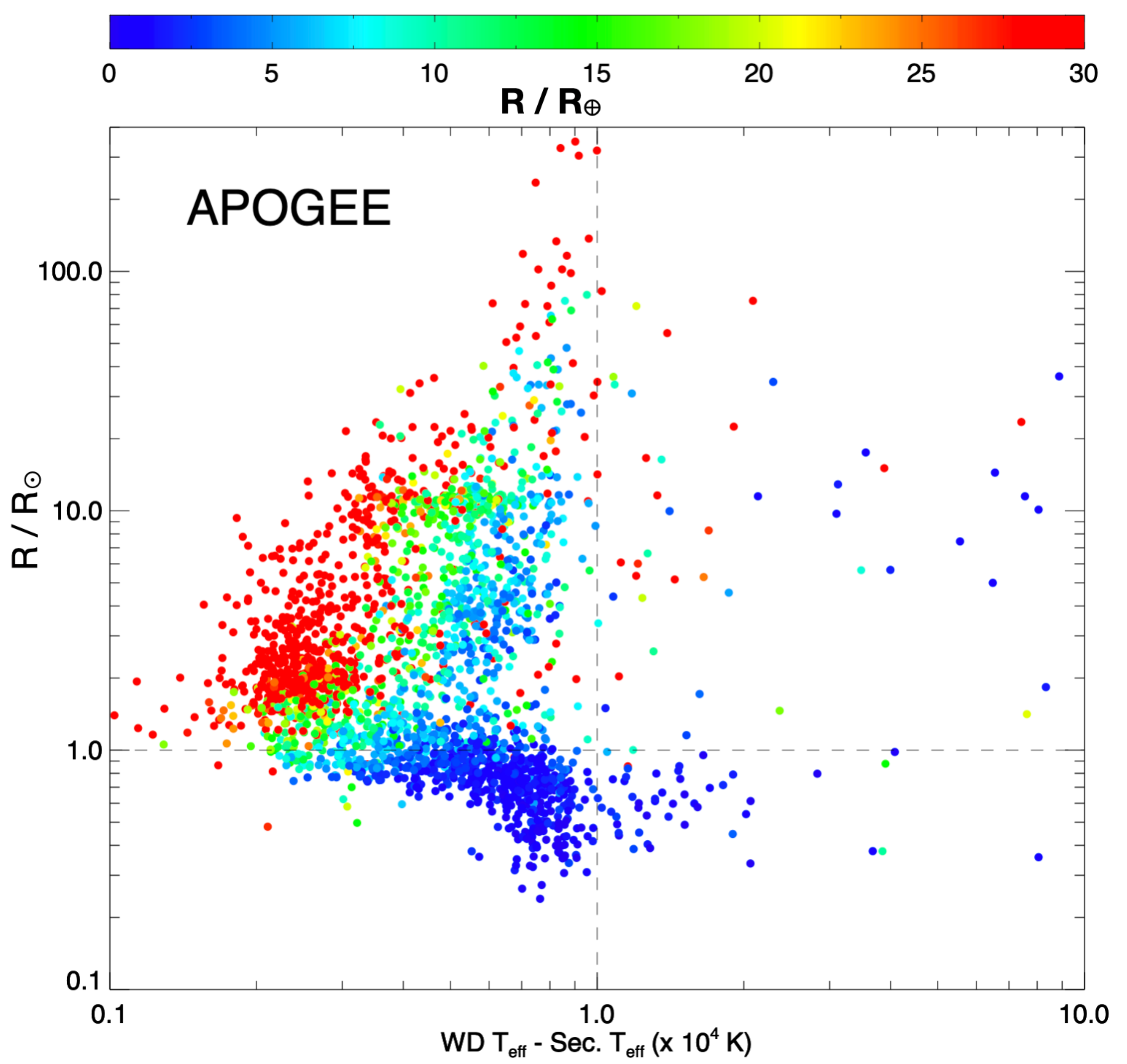}
\caption{\emph{Left panel:} APOGEE-derived  effective  temperature  versus  UV color for the AGGC sample. The figure is color-code by the estimated WD radii. The vast majority of objects with (FUV - NUV)$_{\circ}$ $<$ 3.5 in the AGGC show a potential WD where the radii is R $<$ 10 R$_{\Earth}$. \emph{Right panel:} The temperature-radius diagram, given by the derived radius of the secondary as a function of the difference between the WD effective temperature and the secondary. The figure is color-coded by the derived WD candidate radius.}
\label{radii_TEFF}
\end{figure*}


Practically, for the blue side of the SED, the constraints for our $\chi^2$ approach are the effective temperature and the radius of the WD.  We use only the models with $\log{g} = 8.5$ from the Koester WD models \citep{Koester2010}, because the WD surface gravity information encoded in the two GALEX bands is negligible. The effective temperature coverage in the Koester models goes from 5,000 K to 80,000 K, but the grids do not always have a uniform coverage of all of parameter space.  In this case, the temperature steps are only 250 K from 5,000 to 20,000 K, then 1,000 K steps are used from 20,000 K to 40,000 K, and then, finally, 10,000 K steps are implemented from 40,000 K to 80,000 K. However, as pointed out earlier, because of the saturation of the ($FUV-NUV$) color (Fig.~\ref{GALEX_WDMS_ACTIVE}), for WDs hotter than 30,000 K we cannot retrieve reliable temperatures from the UV GALEX bands anyway; fortunately, however, the vast majority of WDs have temperatures cooler than 30,000 K (see Fig.~\ref{WD_temp}).

Just as for the secondary star where the radius is given primarily by 2MASS and WISE bands, we use only the $FUV$ and $NUV$ GALEX bands to estimate the WD radius because the contribution in the UV flux from the secondary's photosphere is negligible.
For each given temperature in the WD model grid, we test different radii from 0.1 to 500 R$_{\earth}$ in steps of 0.1 R$_{\earth}$, following Equation~\ref{radius}.   Estimated stellar distances, $r$, came from \emph{Gaia} eDR3 parallaxes and the Bayesian isochrone-fitting code StarHorse \citep{Santiago2016,Queiroz2020}.

Figure~\ref{SED_chi} shows a representative final SED fitting, together with the $\chi^{2}$ surface in the two dimensional parameter space of WD temperature and radius, for a confirmed WDMS binary.\footnote{SDSS J232217.42-005725.5 \citep{Eisenstein2006}} The SED in the upper panel shows the best combined total fit (black line) to the observed fluxes (black dots) using the Kurucz model for the secondary (red line) and the Koester model for the WD (blue line).\footnote{The original models have been rescaled to 4$\pi\times$ Eddington flux in units of erg/cm$^{2}$/s/\AA.} 
The bottom panel shows the $\chi^{2}$ distribution as a function of WD temperature and radius.
The ridge of blue color in this plot represents the valley of low $\chi^{2}$ values, and hence, the WD model that results in the best fits between the model and the data. 
To avoid potential outliers and/or upper limits, the $\chi^{2}$ is calculated using all data points except W3 and W4 from WISE.
The lowest $\chi^{2}$ value for this SED-fitting is represented as the black star symbol, and for this particular system indicates that this occurs for $T_{\rm eff}$ $\sim$ 18,000 K and $R$ $\sim$ 1.0~R$_{\earth}$. 

The radius uncertainties for a given band, $\epsilon$R$_{\lambda}$, are determined by error propagation of Equation~\ref{radius}; that is  
\begin{equation}
    \epsilon R_{\lambda}/R = \left[(\epsilon F_{\lambda}/2F_{\lambda})^2 + (\epsilon r/r)^2\right]^{1/2}
\end{equation}
\noindent where we have assumed that the uncertainties associated with the models are negligible, and so set $\epsilon F_{\lambda, surf}$ = 0. Thus, the radius uncertainties depend on the uncertainties in the observed fluxes, $\epsilon F_{\lambda}$, and the uncertainties in the stellar distances, $\epsilon r$. Like we did for the stellar radius, we use 2MASS and WISE bands for the secondary and FUV and NUV GALEX bands to estimate the uncertainties for the WD. We estimate $\epsilon$R using the average $\epsilon$R$_{\lambda}$ for the bands used in each stellar component. 
To obtain the temperature uncertainties, we first considered to use the radius uncertainties and the $\chi^2$ value: From the $R$ $\pm$ $\Delta R$ values in the temperature-radius plane (see bottom panel in Fig.~\ref{SED_chi}) we select the lowest $\chi^2$ values for the corresponding radii as the lower and the upper temperature values. 
However, the resulting uncertainties were in our assessment overly optimistic; hence we instead adopt 
the comparison between our SED estimates and the SDSS spectroscopic values, for which we find $\sigma_T$ $\sim$ 2200~K.

\subsection{Validation Against SDSS WDMS Binaries}

We tested the WD effective temperatures and radii derived using this SED fitting procedure against these same parameters as derived independently for the SDSS WDMS pairs discussed in \S\ref{wdms_sdss}. To do so, for the SDSS WDMS systems we use the stellar distances derived from \emph{Gaia} eDR3 parallaxes with the probabilistic approach that uses a prior constructed from a three-dimensional model of our Galaxy developed by \cite{Bailer_Jones2021}. In the SDSS catalog, WD parameters like $T_{\rm eff}$ and $\log{g}$ are estimated using the best fitting template and the Balmer lines in the SDSS optical spectra \citep{Schreiber2008,Nebot2009,RM2012}. They also estimated the mass and the radius of the WD by using theoretical cooling models \citep[e.g.,][]{Bergeron1995}, together with the stellar parameters determined from the best line profile fit ($T_{\rm eff}$, $\log{g}$). Because, obviously, $\log{g}$ depends both on mass and on stellar radius, the SDSS WD masses and radii must rely on a theoretical mass-radius relation \citep[e.g.,][]{Provencal1998}. For the comparison exercise, we need to bear in mind that while our radius estimation for the confirmed SDSS WDMS relies mainly on distances and in the ratio between the apparent and absolute flux, the listed radius in the literature rely on the estimation of $\log{g}$ from the Balmer lines in the SDSS spectra, and the theoretical mass-radius relation for WDs. Hence, with this validation exercise we also compared two different approaches to estimate the WD radius in these binary systems.   

With the temperatures and radii for the SDSS WDs in hand, we can compare them to the same values derived from the SED analysis. The left panel of Figure~\ref{SED_sdss} shows the histogram of discrepancies between the WD temperatures derived using the SED analysis and the WD temperature estimated from the SDSS WDMS spectra \citep[e.g.,][]{Schreiber2008,Nebot2009,RM2012}. Fitting a simple Gaussian to the histogram of discrepancies we find a small offset of about $\mu_{T_{\rm eff}}$ $\sim$ $-200$ K and a spread of $\sigma_{T_{\rm eff}}$ $\sim$ 2200 K (see Fig.~\ref{SED_sdss}). Similarly, in the right panel of Figure~\ref{SED_sdss}, we fit the distribution of discrepancies between the WD radii (in Earth radii) calculated for the SDSS optical spectra as described above and those found from the SED analysis. Here we find no significant offset and a spread of $\sigma_R$ $\sim$ 0.6 R$_{\earth}$ from the Gaussian fitting. Thus, we see from these overall good agreements of our calculated WD $T_{\rm eff}$ and radius values with those previously found by other surveys that our SED fitting procedure is reliable, at least for the WDs in WDMS binaries of the type found in the SDSS catalog.


\section{Results: A Comprehensive Catalog of Compact Binaries with White Dwarfs}
\label{sec:results}

In this Section we present the APOGEE-GALEX-\emph{Gaia} catalog to study white dwarfs in close binaries. Table~\ref{tab:catalog} represents the first four lines of our WD binaries sample. Together with the APOGEE and \emph{Gaia} EDR3 IDs, the T$_{\rm eff}$, log g and radius for the secondary, the table shows the estimated WD T$_{\rm eff}$ and radii in R$_{\earth}$. In the next Section, we discuss some physical features of the sample like the WD T$_{\rm eff}$, and radii. 


\begin{deluxetable*}{rcccccccc}
\centering
\tablewidth{0pt}
\tabletypesize{\footnotesize}
\tablecaption{WD binaries APOGEE DR17 catalog \label{tab:catalog}}
\tablecolumns{9}
\tablehead{
\colhead{APOGEE ID} & \colhead{\emph{Gaia} EDR3 ID} & \colhead{RA (J2000)} & \colhead{DEC (J2000)} & \colhead{WD T$_{\rm eff}$}& \colhead{Sec T$_{\rm eff}$} & \colhead{Sec log g} & \colhead{R$_{\rm WD}$} & \colhead{R$_{\rm sec}$} \\
\colhead{} & \colhead{} & \colhead{(degrees)} & \colhead{(degrees)} &  \colhead{(K)} & \colhead{(K)} & \colhead{(cgs)} & \colhead{(R$_{\earth}$)} & \colhead{(R$_{\odot}$)}}
\startdata
2M00001362-1913042 &  2413936998069050496  & 0.0568 & -19.2178 & 10683 & 5555 & 4.3 & 5.2 & 1.1 \\
2M00031637+0203553 & 2739046437325768704   & 0.8182 & 2.0653 & 10656 & 4747 & 2.9 & 9.7 & 6.7 \\
2M00042113+0109145 &  2738372917734134144  & 1.0881 & 1.1540 & 11810 & 4838 & 3.4 & 2.9 &  3.1 \\
2M00081185-5220420 & 4972421528506663552   & 2.0494 & -52.3450 & 9796 & 3632 & 4.7 & 0.3 & 0.4   
\enddata
\end{deluxetable*}

\subsection{Radius and Temperature Distributions}

Figure~\ref{radius_distribution} shows the derived stellar radius distributions for the MS (black line) and RG (red dashed line) secondary stars (left panel, for which the units shown are solar radii) and those for the WD candidate primaries (right panel, for which the units shown are in Earth radii). We broke the distribution between MS and RG using the APOGEE surface gravity, where we use a log g $>$ 4 to select MS and log g $<$ 4 for the subgiants and RG.     
The distribution of MS secondary stars shows a clear peak
near about one solar radius, and range from 0.1 to 3.5 R$_{\odot}$. While the RG we have very clear peak around $R$ $\sim$ 10 R$_{\odot}$, which is dominated by red clump (RC) giants,
the radius for the subgiants and RG show a range from 1.5 to $\sim$ 300 R$_{\odot}$ 
(see also Figures~\ref{WD_radii_MS} and \ref{radii_TEFF}). In the right panel of Figure~\ref{radius_distribution} we also have the derived stellar radius distributions for the WDs. The two histograms represent the MS (black line) and RG (red dashed line) secondaries, respectively. Interestingly, the number of WDs with a R $<$ 3.5 R$_{\earth}$ is much larger for the MS objects than for the RG. The WD stellar radius distribution for the RG shows also another peak where the radius of the WD candidate is 100 R$_{\earth}$. These objects will be removed from our final WD binary sample (see also Fig.~\ref{WD_radii_MS}). The vertical line in the right panel of Figure~\ref{radius_distribution} represents $R = 25 $R$_{\earth}$, an upper limit for our WD binary sample.  

Figure~\ref{UV_color_TEFF} shows the WD tempertures and radii against the GALEX UV colors, for WD radii up to 20~R$_\Earth$. The WD temperature derived using the individual SEDs agrees with the expected UV color for the individual WDs, as we discussed for SDSS WDs in Figure~\ref{WD_temp}. Moreover, as shown by the color-coding for calculated radius of the candidate WD in the system,
the population of systems with WDs hotter than 10$^{4}$ K are dominated by those with radius smaller than 5 R$_{\Earth}$, whereas the population with temperatures cooler than 10$^{4}$ K are dominated by objects with a larger stellar radius than the expected radius range for a WD. 
However, WDs with radius around 20 R$_{\earth}$ have been reported for WD binaries systems \citep[e.g.,][]{Sokoloski2006,Lewis2020}; such an inflated radius for a WD can indicate the presence of a disk around the star.  

We use the derived photometric stellar radii to build the MS/RG radii versus WD radii diagram shown in Figure~\ref{WD_radii_MS}. This diagram can be used to refine our selection of WD binary candidates. 
The radii derived using GALEX bands appears in Earth radius units, while the radii using IR-bands is in solar units.
The top panel of Figure~\ref{WD_radii_MS} shows the the sample of {\it bona fide} WDMS binaries found in the SDSS (\S\ref{wdms_sdss}, Fig.~\ref{GALEX_WDMS_ACTIVE}). The number of SDSS WDMS pairs with $T_{\rm eff}$ $>$ 4500 K drops suddenly because the flux from the secondary can dominate the optical spectrum, obscuring the flux from the WD.
This bias should be less severe in the present exercise, where we use the GALEX UV bands to identify the WD. For instance, we find WD binaries where the primary can be hotter than $\sim$ 4500 K (see black points in Fig.~\ref{GALEX_APOGEE}). The SDSS WDMS sample is a useful guide to 
where some confirmed WD binaries should appear in the radius-radius diagram. We highlight the limit to where these objects live using a red line in the figure. This population is dominated by stars with $R$ $<$ 0.8 R$_{\odot}$. 

In the middle panel of Figure~\ref{WD_radii_MS} we show the AGGC sample for those objects for which we have calculated the radii using SEDs and APOGEE DR17 StarHorse distances along with \emph{Gaia} eDR3 parallaxes. The AGGC sample 
clearly consists of systems with main-sequence secondaries with $R$ $<$ 2 R$_{\odot}$ as well as with sub-giants and giant star secondaries with $R$ $>$ 2 R$_{\odot}$. We also see a clear signature of an RC population, as the spike of systems visible at $R$ $\sim$ 10 R$_{\odot}$. 

Finally, we also show in Figure~\ref{WD_radii_MS} (lower panel) the locus of \emph{chromospherically active stars} from the \cite{Boro2018} catalog discussed in \S\ref{CA}. We found that stars in the ``cool dwarf branch" going from $\sim$5000 K to $\sim$3900 K discussed in Figure~\ref{GALEX_WDMS_ACTIVE} show a radius derived from the UV bands with $R$ $<$ 10 R$_{\Earth}$. Thus, while the chromospherically active sources primarily mimic single stars in Figure~\ref{WD_radii_MS}, some also resemble WD binaries. However, such stars may not be ``contaminants'',
but systems for which chromospheric activity itself has been incited by interaction with a close WD companion \citep[e.g.,][]{Bleach2002}.


\subsection{Temperature-Radius Diagram}

We investigate the APOGEE-derived effective temperature of the secondary star as a function of UV color in the left panel of Figure~\ref{radii_TEFF}; the systems represented in this figure are color-coded by the derived stellar radius for the WD candidate. 
Meanwhile, the right panel of Figure~\ref{radii_TEFF} shows 
a \emph{temperature-radius diagram}, given by the derived radius of the secondary as a function of the difference between the WD temperature and the secondary.
The points in this figure are also color-coded by the derived WD radius.

Together, the two panels of Figure~\ref{radii_TEFF} reveal that the the majority of systems with $R$ $>$ 25 R$_{\Earth}$ for the potential WD have ($FUV - NUV$)$_{\circ}$ $>$ 3.5, a color value that corresponds to a WD $T_{\rm eff}$ $\leq$ 9000 K (see also Fig.~\ref{WD_temp}). 
This suggests that WD binary candidates with the WD $T_{\rm eff}$ $\leq$ 9000 K are very likely dominated by a non WD companion. By contrast, the number of WD detected in SDSS with $T_{\rm eff}$ $<$ 10$^{4}$ quickly drops \citep[e.g.,][and also Fig.~\ref{WD_temp} in this work]{Kepler2016}, suggesting that for ($FUV - NUV$)$_{\circ}$ $>$ 3.5 the number of non WD binaries should be significant, as is suggested, in any case, by the large radii found in this exercise. 
As may be seen in the right panel of Fig.~\ref{radii_TEFF},
nearly all the stars with $R$ $<$ 1 R$_{\odot}$ present a $R$ $<$ 10 R$_{\Earth}$ for the WD candidate.  Also, objects where the temperature discrepancies are smaller than 3,500 K are dominated for objects with a radii larger than 25 R$_{\Earth}$.  

\subsection{Final Sample}

These radius-radius and radius-temperature diagrams discussed above provide a useful guide for creating a relatively clean sample of WD binary candidates within the AGGC sample. The red line shown in several panels of Figure~\ref{WD_radii_MS} represents our chosen criterion to select the highest probability WD binaries in the AGGC sample, mindful of the loci traced by the SDSS WDMS binaries and that traced by the presumed single stars. A total of 1,806 AGGC stars fall below the red line, and constitute this more reliable sample of WD binary candidates.



\begin{figure*}[ht]
\begin{center}
\includegraphics[width=\hsize,angle=0]{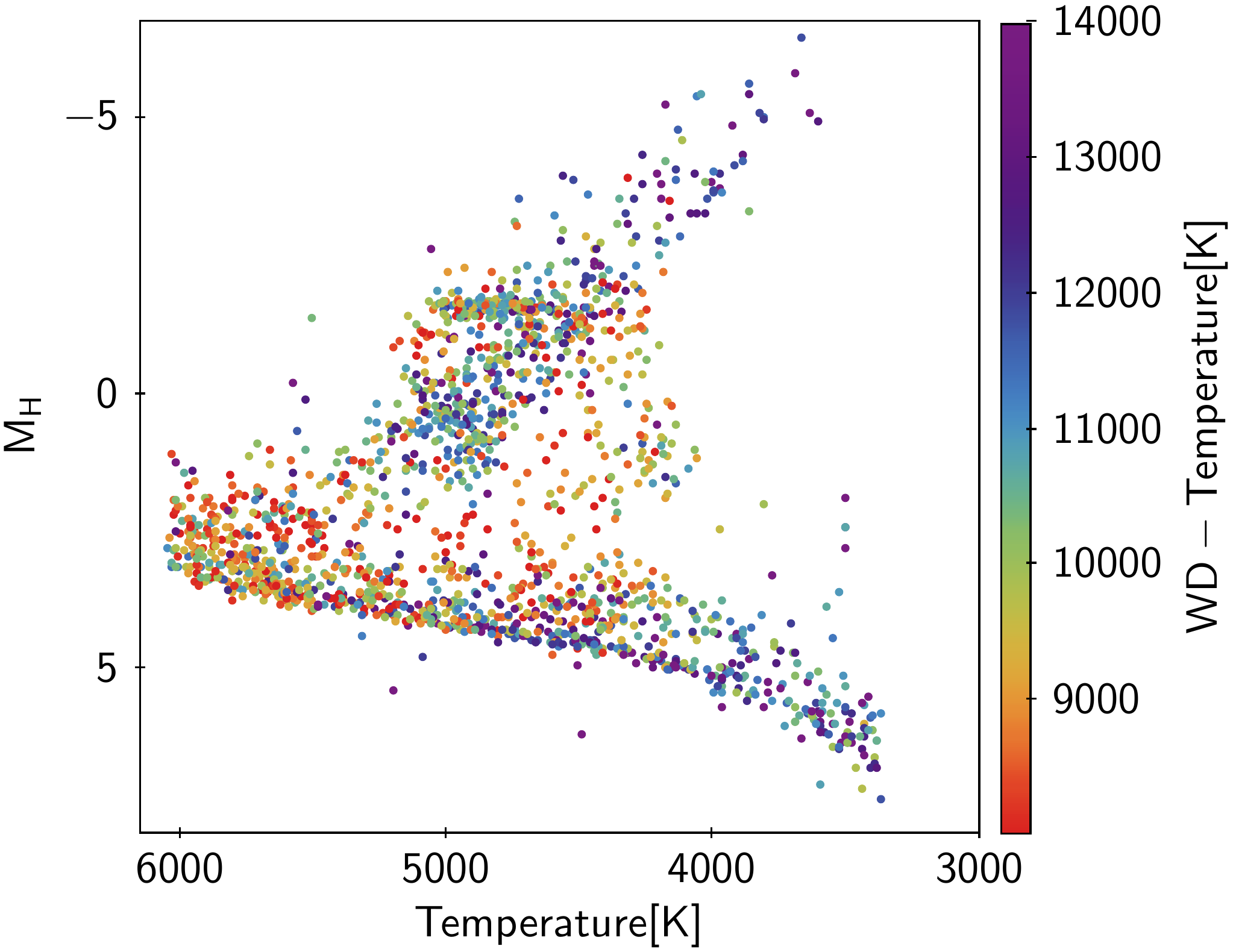}
\end{center}\caption{HR-Diagram for the  APOGEE--GALEX--\textit{Gaia} Catalog, with APOGEE-derived temperatures and $H$-band luminosities from 2MASS photometry$+${\it Gaia} parallaxes.  Sources are color-coded by the inferred WD temperature from the SED-fitting (\S\,\ref{SED_sec}).}
\label{CMD_WDMS}
\end{figure*}

\section{Discussion: Color-Magnitude Diagram and Compact Binary Evolution}
\label{CMD_sect}


\begin{figure*}[ht]
\includegraphics[width=\hsize,angle=0]{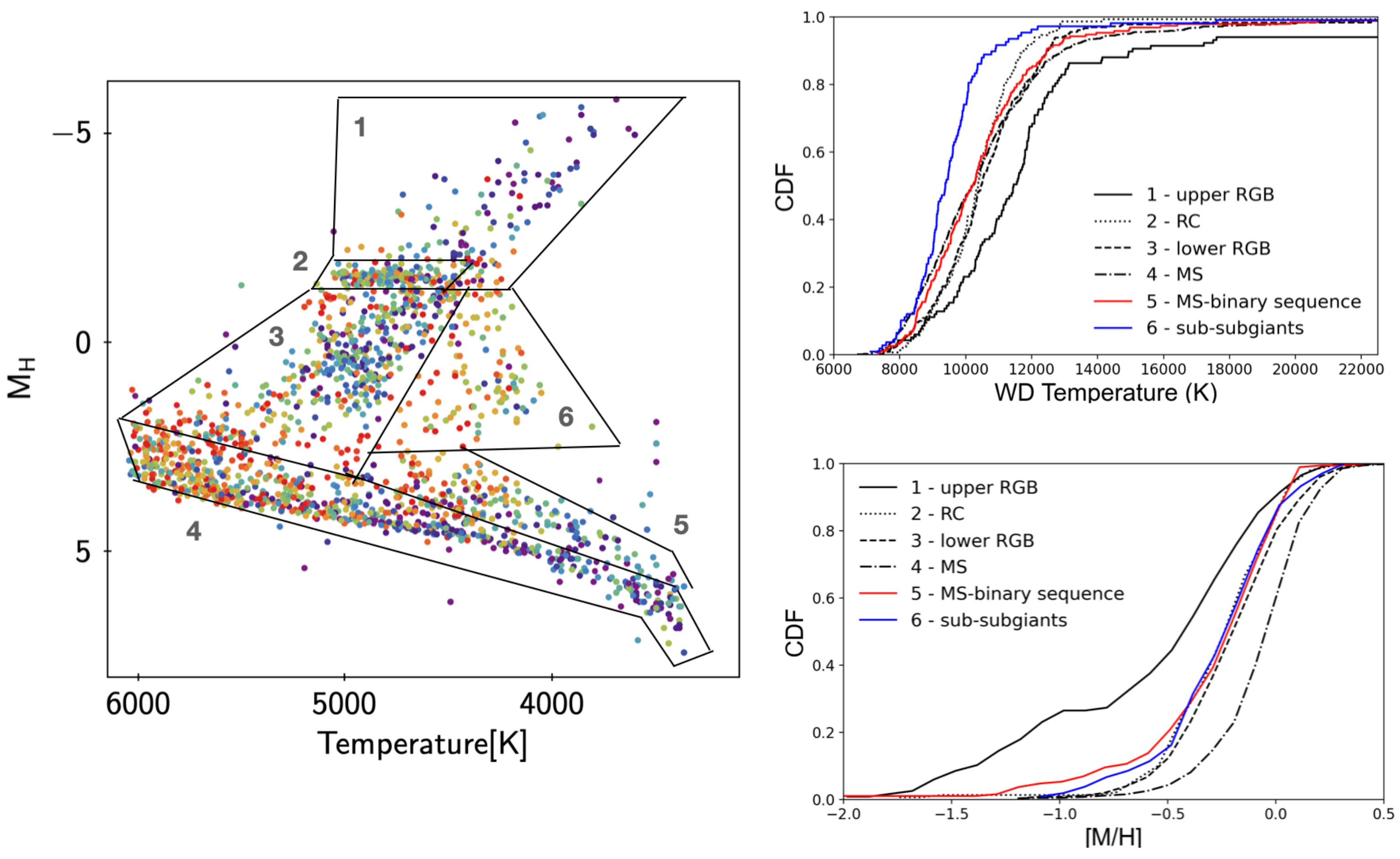}
\caption{\emph{Left panel:} Same as Fig~\ref{CMD_WDMS}, but now divided into regions corresponding to principal stellar evolutionary phases: 1 - upper RGB, 2 - RC, 3 - lower RGB, 4 - MS, 5 MS-binary sequence and 6 - sub-subgiants. \emph{Right panel:} The cumulative distribution functions of both the WD temperature (upper panel) and overall metallicity (lower panel) for stars lying in the different regions of the CMD. The upper-RGB shows a larger number of hotter WDs and the metallicity distribution for these systems is clearly skewed to lower metallicities with respect to those in the other regions of the CMD, while the main sequence group of binary candidates shows the largest number of metal-rich objects.}
\label{CDF_temp_meta}
\end{figure*}

As a demonstration of the potential for the AGGC to address numerous scientific questions,
in this Section we do a pilot exploration of the color-magnitude diagram (CMD) for the selected WD binary candidates described above, and investigate how various properties of WD binaries vary as a function of the evolutionary state of the secondary star in the system. To build the CMD we use (a) the effective temperatures from APOGEE DR17 \citep{sdss17}, (b) the StarHorse distances \citep{Queiroz2020} including \emph{Gaia} eDR3 parallaxes \citep{Lindegren2020}, (c) the near-infrared $H$ (1.25 $\mu$m) bandpass apparent magnitudes from 2MASS \citep{Skrutskie2006} to compute the absolute magnitude $M_{H}$ corrected from (d) reddening using the extinctions values provided in the StarHorse catalog. When distances from the StarHorse catalog are not available for our objects we use the photogeometric distances listed in \cite{Bailer_Jones2021}.  


The resulting CMD of the WD binary candidates is shown in Figure~\ref{CMD_WDMS}, which is color-coded by the inferred effective temperature of the potential WD in these pairs (see \S\ref{SED_sec}). The figure shows a well defined main-sequence, a good number of sub-giants, and also a well defined red giant branch together with a prominent red clump (RC). Some WD binaries candidates appear in the binary sequence above the MS; in the case of our WD companion population, these would likely be triple-star systems comprising the visible APOGEE MS star, the WD companion, and another luminous companion comparable in brightness to the APOGEE MS star. Finally, we note a number of systems populating the space between the MS and the sub-giant branch, with $T_{\rm eff} \sim 4500$~K and $M_H \sim 1$. These would appear to be so-called sub-subgiants; we return to discuss this interesting population in \S\ref{Proper_HR}.

A sample of highly likely WD binaries identified across the CMD is an important step toward furthering our understanding of compact binary evolution. One example where improvement is possible (already discussed in Sec.~\ref{SED_sec}) is that SDSS WDMS identified using optical spectra alone are less able to discern bimodal spectral energy distributions (SEDs) when the MS star is hot, and thus such a resulting WD binary survey is strongly biased to those with very late type companions. Furthermore, our knowledge of the fundamental statistics of stellar multiplicity, e.g., multiplicity fraction, period distribution, is still poorly understood, especially for evolutionary stages after the MS and for a volume of study larger than the solar neighborhood \citep[e.g.,][and references therein]{badenes2018}. In the following Section we analyze some properties of our binary sample across the HR diagram in an initial, pilot assessment of this very rich database.

\begin{figure}[ht]
\includegraphics[width=\hsize,angle=0,trim=30 10 10 20,clip]{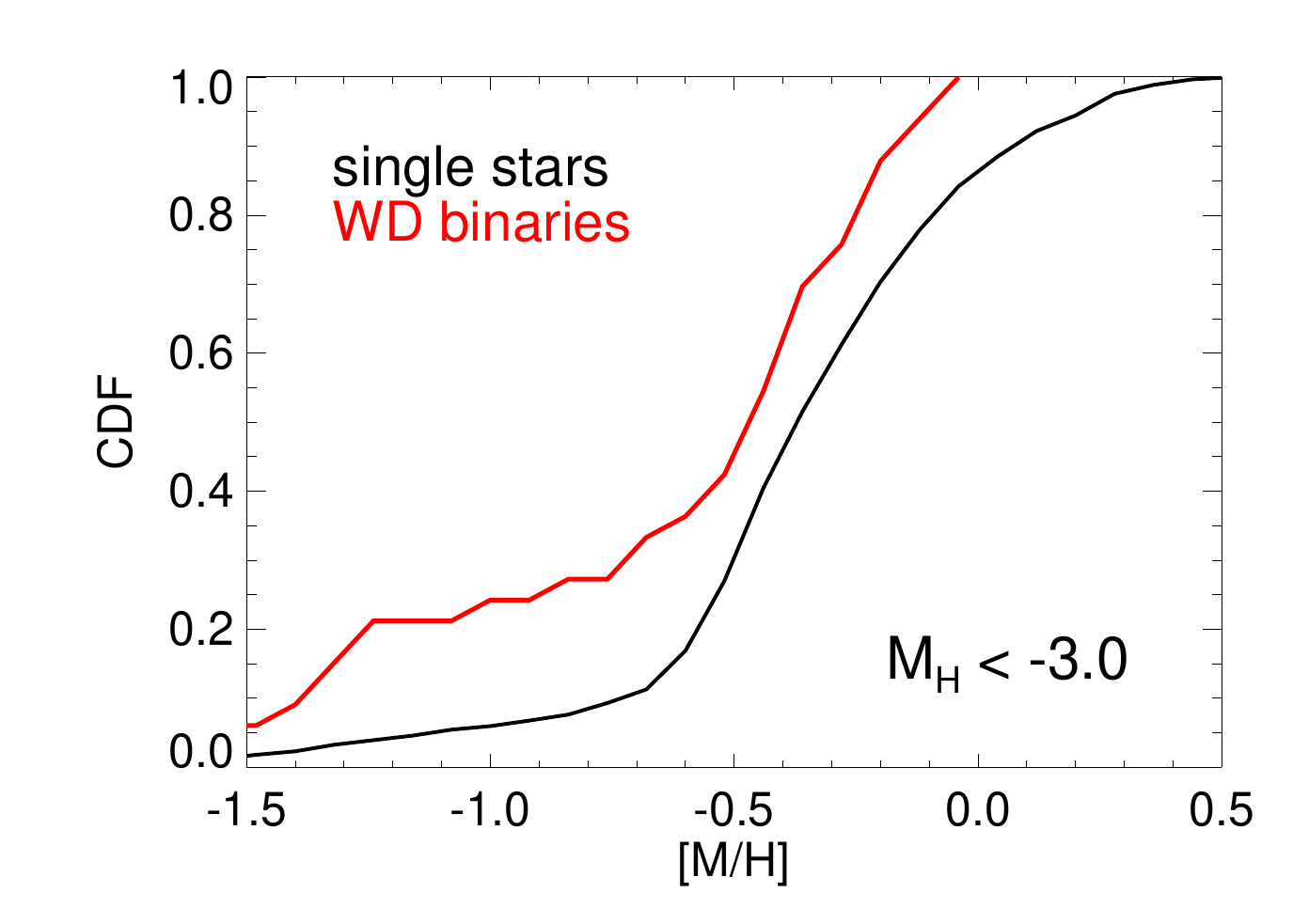}
\caption{Overall metallicity from the APOGEE spectra cumulative distribution function for objects with M$_{\rm the H}$ $<$ -3.0. The black line is for a sample dominated by single stars, the red line for the WD binaries sample discussed in this work. For the most luminous stars the MDF is clearly different for the binary and the non-binary samples.}
\label{CMD_WD_binaries_Lumi}
\end{figure}

\subsection{Properties of White Dwarf Compact Binaries Across the HR Diagram}
\label{Proper_HR}

We first examine and compare properties like the WD temperature and the overall metallicity of the secondary in different regions of the CMD. Figure~\ref{CDF_temp_meta} shows the WD binaries CMD presented in Figure~\ref{CMD_WDMS} but now broken up into areas 
corresponding to WB binary systems containing secondaries on the upper RGB, the RC, the lower RGB and subgiants, the MS, the MS-binary sequence, and the sub-subgiants. 
On the right of Figure~\ref{CDF_temp_meta} we have two panels. The upper panel shows the CDF for the WD temperature, while the lower one shows the CDF for the APOGEE overall metallicity. We find that the upper RGB shows the largest amount of hottest WDs, followed by the MS population, compared to the other regions of the CMD. Interestingly, the RC temperature distribution shows a WD temperature distribution closer to the lower RGB region. The MS-binary sequence suggest that these systems are triplets where one component is a WD; its WD temperature distribution is similar to the MS sample.
The sub-subgiant CDF also shows a different distribution with respect to the other populations. In the lower panel we have the CDF for the metallicity. We find that the metallicity distribution function for the upper RGB (black solid line) differs from the rest of metallicity distributions from different populations in the CMD. 

The number of metal-poor systems ([Fe/H] $<$ -0.7) on the upper RGB is much larger than for the RC and the lower RGB. 
Such a metal-poor tail for the upper RGB compared to other regions in the CMD cannot just simply be explained because the volume of study is larger and there are a larger number of halo objects. 
To understand better the discrepancies we find for the overall metallicity of the WD binaries on the upper RGB and to gain insights into the potential abundance variations induced in the secondary star during the CE phase, it is worthwhile to investigate the differences across the CMD in the metallicity distribution function between systems dominated by WD binaries and a control sample dominated by single stars.

\begin{figure}[ht]
\includegraphics[width=\linewidth,angle=0,trim=0 10 40 50,clip]{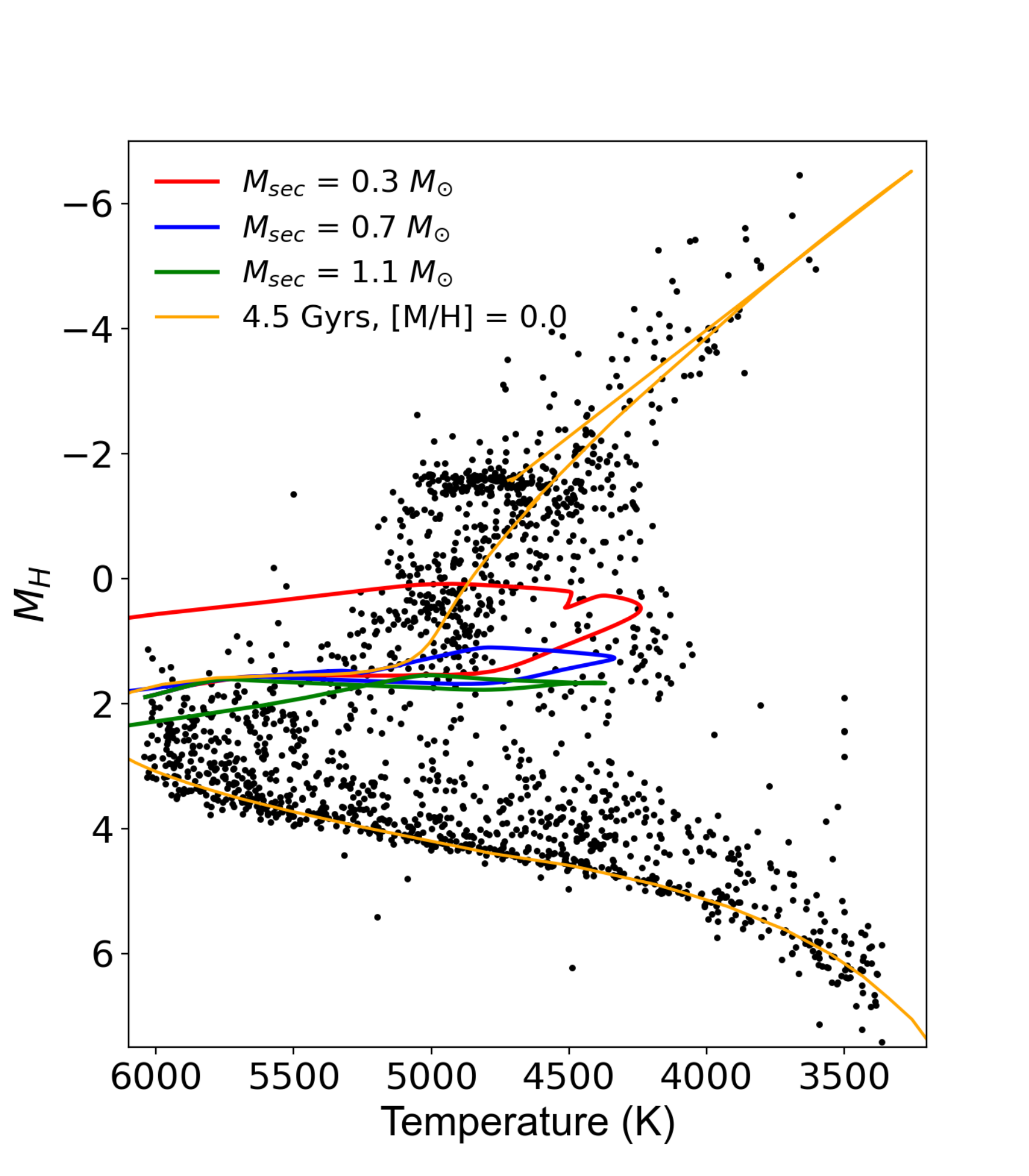}
\caption{Exploring potential sub-subgiant systems in our catalog. Superposed are evolutionary models from \citet{Leiner2017} of sub-subgiants experiencing mass transfer. All models assume a 1.3 M$_{\odot}$ subgiant primary with a compact companion in a 1.0~day orbit. Mass of the companion for each system is indicated by color of the track: 1.1 M$_{\odot}$ (green), 0.7 M$_{\odot}$ (blue), and 0.3 M$_{\odot}$ (red). The \citet{Leiner2017} models have been arbitrarily shifted 200~K cooler to better match the observed span of the SSG+WD candidate systems we observe, suggesting that similar evolutionary configurations but with different initial conditions (e.g., different subgiant mass, etc) could potentially explain the diversity of SSG+WD systems we have identified. We also show in yellow a 4.5 Gyrs solar metallicity isochrone \citep{Bressan2012}}
\label{sub_models}
\end{figure}

To build a closely matching, single-star control sample we subsampled the APOGEE database using
the following flags and selection criteria
\begin{verbatim}
NVISIT > 5 
VSCATTER < 300 m/s
BAD_PIXELS
VERY_BRIGHT_NEIGHBOR
LOW_SNR
STAR_BAD
\end{verbatim}
This gives a total of 12,443 APOGEE targets that should be
dominated by single stars. 
A crossed-match of this ``single-star'' catalog with the double-lined spectroscopic binary sample identified by \cite{Kounkel2021} yields only 26 stars in common, which we remove. In addition, because we want to build a sample of stars that is as random and unbiased as possible, whereas the APOGEE survey has a number of focused science programs that target specific classes of objects \citep{Beaton2021,Santana2021},
we remove targets in fields
associated with these special programs.
This, for example, removed from consideration the stars from a number of Milky Way satellite galaxies targeted by APOGEE, which, of course, have different chemical evolution histories than the Milky Way \citep[e.g.,][]{Tolstoy2009,Hasselquist2021}. 
A comparison of the metallicity distribution function for this ``single-star'' control sample to that for the WD binary sample when limited to the most luminous objects ($M_{\rm H} < -3.0$) shows clear differences (Fig.~\ref{CMD_WD_binaries_Lumi}), with the WD binary sample tending to be more metal-poor than the sample dominated by single stars. 
This may be a product of the strong anti-correlation between close binary fraction and chemical composition \citep{mazzola2020}.
The advantage of this comparison is that this metallicity difference is not driven by variations in sample volumes, since both the single-star and WD-binary samples are drawn similarly from the same parent sample. 

Meanwhile, at the low luminosity end of the HR-diagram, there is a group of systems in Figure~\ref{CDF_temp_meta} outside of the selected areas in the CMD with temperatures lower than 4000 K. A careful look reveals that a few of them have large uncertainties in their parallaxes making their distances, and hence their position in the CMD, less reliable. We checked these and confirmed that they are young stellar objects in known star forming regions. 
However, the ones close to the MS-binary sequence (Group 5) shows very large values for the \emph{Gaia} parameter Renormalized Unit Weight Error (RUWE), suggesting that they are multiple systems \citep[e.g.,][]{Belo2020,Keivan21} potentially containing a WD. 

Finally, we highlight the existence of what appears to be a prominent sub-subgiant population within the WD binary sample (represented by region 6 in Fig.~\ref{CDF_temp_meta}).
Sub-subgiant (SSG) stars have been recognized as likely representing unusual stellar evolution pathways ever since their initial detection as anomalies in the CMDs of some open clusters \citep[see, e.g.,][and references therein]{Mathieu2003}. Subsequent studies of SSGs in clusters have proffered several possible interpretations for these systems: mass transfer in a binary system, collision of two MS stars, mass loss of subgiant envelopes through dynamical encounters, and reduced luminosity due to the strong surface coverage of magnetic starspots \citep[see, e.g.,][]{Leiner2017}. Some recent works have concluded that mass transfer and dynamical formation pathways are disfavored based on the small numbers of SSGs in open clusters, preferring instead the strong starspot interpretation \citep[e.g.,][]{Gosnell2022}. However, attempts to identify and characterize the broader SSG population in the field have only very recently begun \citep{Leiner2022}. Thus, the large population of apparent SSGs in the field identified in Figure~\ref{CDF_temp_meta}, and in particular the knowledge in this work that these SSGs all possess a WD companion, could be an opportunity to make substantial new progress in understanding these enigmatic systems.

For example, one possibility for creating an SSG-WD system could be through a mass transfer channel. This could involve a scenario such as the following: First, start with a MS-MS binary in a wide orbit. When the more massive MS star evolves into a giant, unstable mass transfer occurs so that the system evolves through a common envelope. Given the temperature of the WDs we infer for these systems, the cooling ages suggest this would have occurred on the order of a few 100 Myr ago. Following the common envelope phase the system emerges as a MS-WD binary with an orbital period of a few days or less. Finally, the remaining MS star begins to evolve off the MS, and mass transfer starts again on the subgiant branch. As it loses mass, this second star evolves into the SSG region of the HR diagram, as demonstrated by \citet{Leiner2017}. The specific evolutionary track depends on the mass of the SSG, the mass of the WD, the initial orbital separation, and the fraction of the mass lost by the donor that is accreted by the companion. However, representative tracks from \citet{Leiner2017} are shown in Figure~\ref{sub_models}, and these suggest that at least some of the SSG+WD systems we have identified could plausibly represent such a scenario. Assuming different initial conditions, such as alternative subgiant star masses, could potentially shift the tracks in Figure~\ref{sub_models} to cover more of the SSG+WD parameter space that we observe.

Additional detailed modeling of these intriguing possibilities will be an exciting direction of exploration for a future analysis leveraging the catalog of WD binaries presented in this paper.

\begin{figure*}[ht]
\includegraphics[width=\linewidth,angle=0]{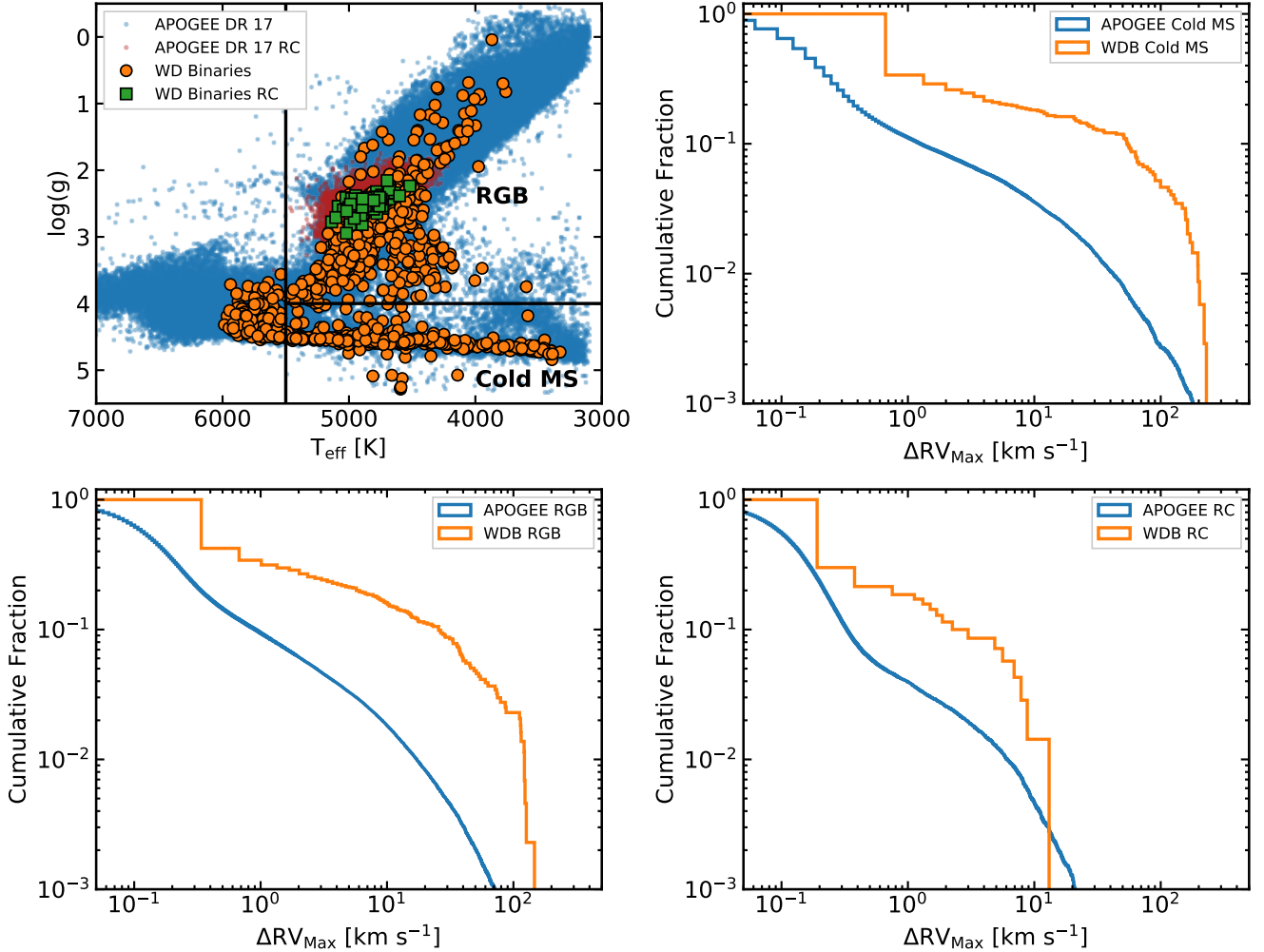}
\caption{\emph{Upper left panel:} $T_{\rm eff}$ - $\log{g}$ diagram for the APOGEE DR17 control sample (blue) and the WD binaries catalog (orange). The selection of cold MS, RGB and RC is also shown. \emph{Upper right panel, lower left and right panels:} $\Delta$RV$_{max}$ cumulative fraction histogram for the cold MS sample, RGB and RC in the APOGEE (blue) and WD binaries catalogs (orange), respectively. The cumulative fraction for the WD binaries is skewed towards larger $\Delta$RV$_{max}$ with respect to the APOGEE control sample for the three selected populations.}
\label{WDB_cum}
\end{figure*}

\subsection{Binary Properties as a Function of RV Variability and Orbital Period}
\label{time}

An advantage conferred by the APOGEE survey is that multi-epoch data were obtained, and RVs from these multiple epochs enable studies of stellar and substellar multiplicity \citep{Troup16,badenes2018,Lewis22}.\footnote{The detection of stellar multiplicity as evidenced by RV variability was one of the motivations for APOGEE being a multi-epoch survey \citep{2017AJ....154...94M}.}
When a sufficient number of epochs are available for a particular star, it is possible to derive or constrain orbital parameters (e.g., \citealt{PW2020}; see below), however, this only applies to a minority fraction of APOGEE targets, since the more typical number of APOGEE visits is only three per star. However, under such circumstances it is still possible to undertake a statistical assessment of the number of stars with close companions, thereby accessing a majority of the APOGEE sample.

A particularly sensitive parameter for this purpose is
the difference between the highest and lowest measured RVs for each WD binary candidate, $\Delta$RV$_{max}$ = max(RV) - min(RV) \citep[see][for a discussion of this metric]{Badenes2012}.
Figure~\ref{WDB_cum} shows the cumulative histograms for $\Delta$RV$_{max}$ in two samples: (1) 
the full APOGEE DR17 sample (blue solid line) following a similar selection to that described in \cite{mazzola2020}, and (2) the main WD binary sample (orange solid line).  In this case, we divide the analysis into three broad categories: (a) the ``cold MS'' ($T_{\rm eff} < 5500$ K and $\log{g} < 4.0$), (b) the RGB ($T_{\rm eff} < 5500$ K and $\log{g}> 4.0$), and (c) the RC, using the catalog of red clump stars in the APOGEE DR17 sample following \cite{Bovy2014}. We show these divisions in the upper left panel of Figure~\ref{WDB_cum}, where blue points show the APOGEE DR17 reference sample, and orange points show the WD binary catalog. As may be seen in the other three panels of Figure ~\ref{WDB_cum}, the $\Delta$RV$_{max}$ CDF for the WD binaries (orange solid lines) is clearly skewed towards larger $\Delta$RV$_{max}$ values, suggesting shorter periods for these systems. The RC sample shows the largest difference between the WD binary sample and the APOGEE reference sample. These results can be interpreted as evidence for drastic loss of angular momentum associated with the formation of the WD, most naturally explained by a CE episode leading to the ejection of at least some of the envelope of the mass primary/WD progenitor.

The above, statistical analysis of orbital kinematics in the WD binary sample already unlocks tantalizing results worthy of further exploration.  But the WD binary sample drawn from the AGGC has also produced the largest sample of such systems having
uniformly derived orbital parameters.
These systems, which contain secondaries broadly spanning the HR Diagram, are a uniquely powerful tool to be exploited for very detailed analyses of WD binary evolution. To explore this potential, we rely here on 
an APOGEE DR17 value added catalog \citep{PW2020} containing posterior samplings of Keplerian orbital parameters (e.g., orbital period) derived using {\it The Joker} \citep{Price2017} for all DR17 stars having three or more APOGEE RV measurements. For this exercise, we select out a preliminary set of likely binaries, where we select RV-variable sources using a log likelihood ratio comparing {\it The Joker}'s best fit of the APOGEE RVs to a best fit constant velocity model.
Cross-matching our catalog with this VAC yields 252 potential WD binaries having well-constrained orbital periods. More than half of the systems have more than four epochs while around 120 systems have 3 $\leq$ epochs $\leq$ 4. Nearly all the systems show an uncertainty in their periods smaller than one day. 


\begin{figure*}[ht]
\includegraphics[width=\linewidth,angle=0]{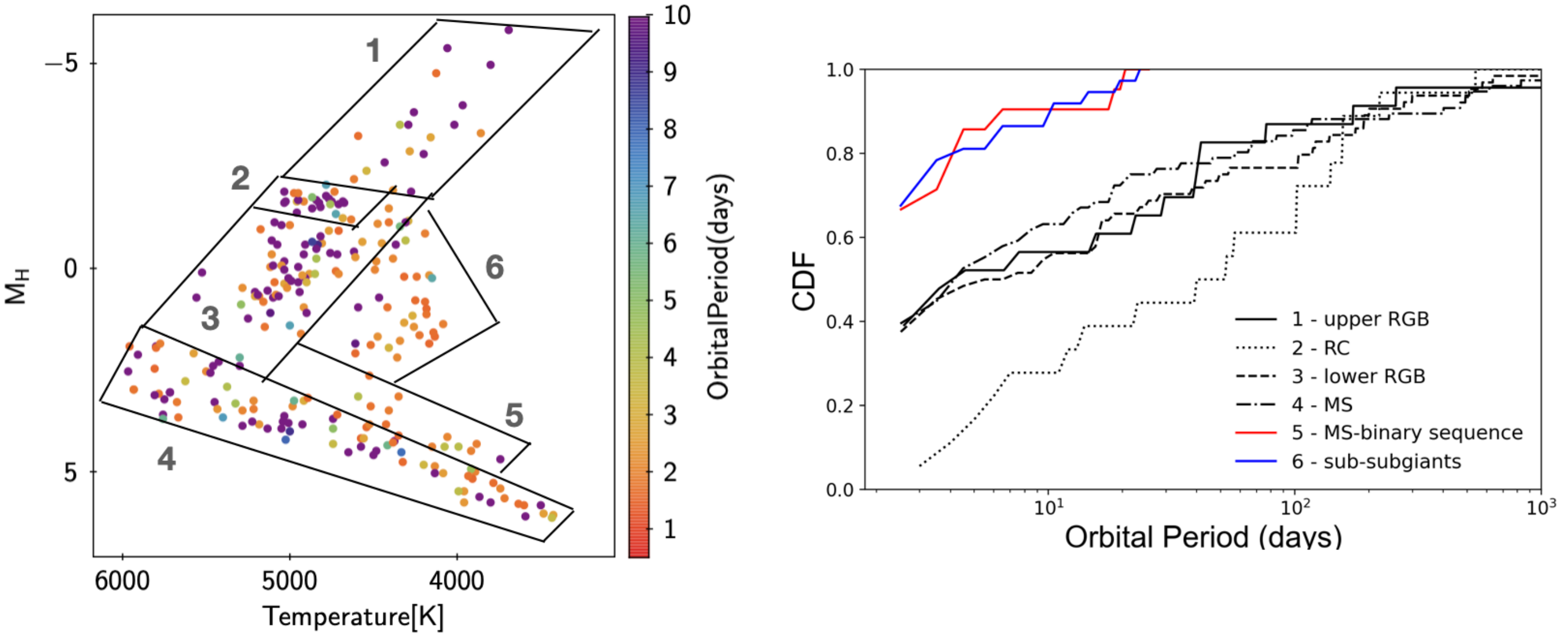}
\caption{\emph{Left panel:} The CMD distribution of WD binaries having reliable, full orbital fits to the APOGEE multi-epoch RVs, color-coded by the derived orbital period. We divide the sample of 252 such systems into secondary star evolutionary stages
as follows: 1 - upper RGB, 2 - RC, 3 - lower RGB, 4 - MS, 5 MS-binary sequence and 6 - sub-subgiants. \emph{Right panel:} Cumulative fraction of the orbital period in days for the different CMD groupings. 
The RC, the MS-binary sequence, and the sub-subgiants have clearly a different period distribution with respect to the other CMD groups.  
The RC systems shows the smallest number of short period binaries, while the MS-binary sequence and the sub-subgiants are greatly skewed to having 
the largest number of short period binaries.}
\label{WDB_P_CDF}
\end{figure*}



Figure~\ref{WDB_P_CDF} shows the HR Diagram of these 252 likely
WD binaries having well-constrained orbital periods, with symbols color-coded by that derived orbital period in days. While the WD binary candidates cover widely the range of sensitivity for the combined, decade-long APOGEE-1 and -2 surveys, 
a large fraction of our systems show periods of only a few days. The right panel of Figure~\ref{WDB_P_CDF} shows the CDF of the orbital period for the 252 sample stars, broken up similarly into the six secondary star evolutionary groups used in \S\ref{Proper_HR} (compare the left panel of Fig.\ref{WDB_P_CDF} with the left panel of Fig.\ref{CDF_temp_meta}).
While the orbital period distributions show a similar behaviour for the upper and lower-RGB and the MS, 
the RC distribution appears to be heavily diminished in the number of short period systems. This likely is the result of the clearing out of residual short period systems during a second CE phase \citep{badenes2018}. 
Meanwhile, the systems lying in the MS-binary and SSG groups are heavily biased toward indicating the presence of a short period system.  In the case of the SSG group, the presence of a close binary is consistent with the subgiant-WD mass transfer formation channel discussed in \S\ref{Proper_HR} and featured in Figure \ref{sub_models}.  In the case of the MS-binary group, which are putative triple systems including a WD, the skew towards short period binaries seen in Fig.\ref{WDB_P_CDF} may reflect evolution of the hierarchical systems due to the Kozai-Lidov mechanism \citep{Thompson11}. However, these systems could be also binaries with low-mass white dwarfs, where the systems are descended from short period main sequence binaries \citep[e.g.,][]{Lagos2022}, since 
such systems are known to very frequently host a distant tertiary \citep{Tokovinin2006}. \cite{Moe2018} have pointed out 
that the occurrence rate of tertiaries to short period binaries cannot be caused by the Kozai-Lidov mechanism alone.




WD binaries with orbital periods of $P < 100$ days are strong PCEB candidates. The cut off at a 100 day orbital period should exclude most binaries that have had no stable mass transfer 
\citep{Nebot2011,Kruckow2021}. Recently, \cite{Lagos2022} suggested that PCEBs are only systems with periods below five days, while systems with periods of the order of several weeks to months could be the result of stable and non-conservative mass transfer. The evolution of PCEBs is driven by angular momentum loss due to gravitational radiation \citep{CF1993} and disrupted magnetic braking \citep{WZ1981,R1983}. 
Subsequent evolution may bring the system into a semidetached configuration. Cataclysmic variables (CVs), composed of a WD as the primary and a low-mass star or a brown dwarf as the secondary, belong to this type of semidetached system, where the secondary fills its critical lobe and transfers mass towards the primary \citep{W2003}. 

\begin{figure}[ht]
\includegraphics[width=1.\hsize,angle=0]{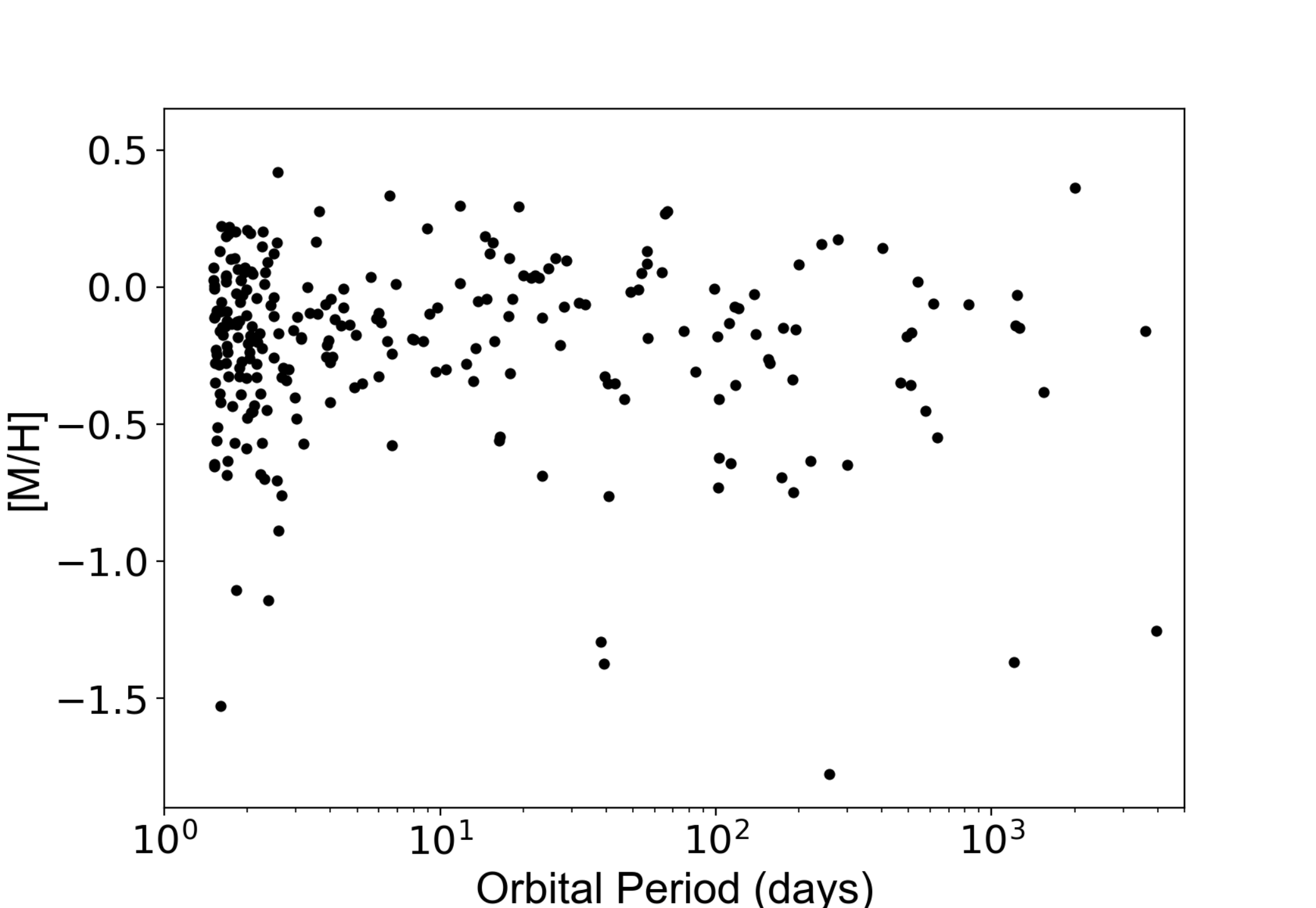}
\includegraphics[width=0.99\hsize,angle=0]{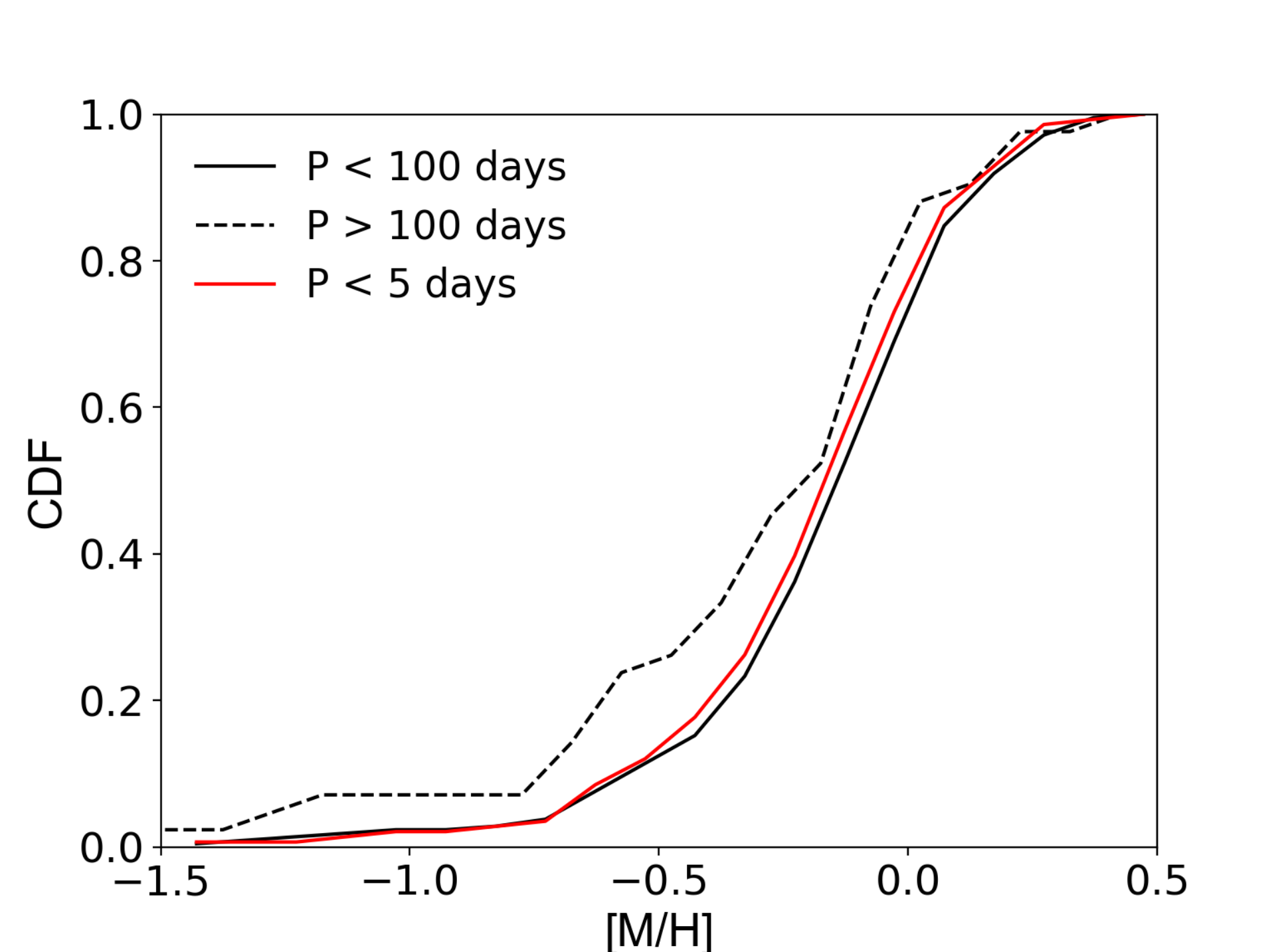}
\caption{\emph{Top panel:} Period-metallicity distribution 
for the 252 WD binaries in our sample with well-defined orbital solutions fit to the APOGEE radial velocities.
The vast majority of these binaries have an orbital period of a few days but span a wide metallicity range ($-0.7<{\rm [M/H]}<+0.3$).  A 
number of metal poor ([M/H]$<-1$) systems are found covering a broad range of periods. \emph{Bottom panel:} The cumulative distribution function of the metallicity for three samples selected according to the orbital period. The 
solid line represent systems with $P < 100$ days and the red line systems with $P < 5$ days, while the 
dashed line shows binaries where $P > 100$ days. Systems with shorter orbital periods tend to have higher metallicity.}
\label{Fe_period}
\end{figure}

How the common-envelope phase and the mass-loss affect binary evolution and the chemical abundances measured for compact binaries is still not well understood \citep[see, e.g.,][and references therein]{mazzola2020}. Motivated by this problem, Figure~\ref{Fe_period} shows the metallicity as a function of the orbital period for the 252 WD binary candidates with well established periods.  As mentioned above (and illustrated in Fig.\ref{WDB_P_CDF}),
a large fraction of these systems show an orbital period of a few days, and this is in good agreement with the $\Delta$RV$_{max}$ exercise performed on the much larger sample of WD binaries having only a few epochs of data discussed even earlier, where the WD binaries clearly have larger $\Delta$RV$_{max}$ than the bulk of the DR17 stars. The results of both of these analyses suggest that their period distributions must be skewed toward shorter periods, presumably as a result of common envelope evolution. Using the CDF for three different samples selected using the orbital period, we find that the sample dominated by wide binaries (``WB'', defined here as $P > 100$ days) is skewed to lower metallicities than the sample dominated by PCEB candidates ($P < 100$ days).We also note that systems with $P < 5$ days show a similar metallicity distribution (red solid line in Fig.~\ref{Fe_period}); 
thus, at least in terms of metallicity distribution, there is no apparent difference with this more strict definition of the likely PCEB systems.  The Figure~\ref{Fe_period} results are in agreement with our previous analysis of the metallicity distribution for 21 WDMS systems separated into their respective classifications as WB or PCE systems \citep{Corcoran2020}. Here we validate that previous, somewhat tentative result derived from a very small fraction of the AGGC-produced WD binary sample with the now much larger sample of WD binaries that more broadly cover the H-R diagram. Both studies therefore suggest that there is some sort of enrichment of the secondary star's surface chemistry during the CE phase.


As one final demonstration of the scientific reach of our new catalog of 3,414 WD binary candidates, we identify within it some
54 metal-poor systems ([M/H] $< -1.0$), among them four with [M/H] $< -2.0$, and one system with [M/H] $= -2.4$, making it the most metal-poor WD binary candidate \citep{Corcoran2020}.  Among the 252 systems having well-defined orbital solutions, Figure~\ref{Fe_period} shows  orbital periods ranging from a few days to a few thousand days for eight systems having [M/H] $<$ -1.0.

\section{Conclusions}
\label{conclusion}


A majority of MS stars exist in binary systems \citep[e.g.,][and references therein]{Jas1970,DK2013,PW2020}, with $\sim$70 per cent of them predicted to interact with their companion(s) during their lifetime, even higher
in the massive regime \citep{Kobul2014}. It is not possible to develop a comprehensive theory of stellar evolution without taking into account stellar multiplicity
\citep[e.g.,][]{DeMarco2017}. In this study we focused on potential WD binaries found within the APOGEE survey using the APOGEE-GALEX-\emph{Gaia} catalog (\S\ref{catalog}), and showed how the use of SEDs including UV-band photometry is an efficient method to identify such objects \citep[e.g.,][]{Morgan2012,Parsons2016}. 

One of the least understood phases of compact binary evolution is 
the CE phase \citep{Paczy1976}, where one of the main issues is our ignorance of the efficiency of the energy transfer between the orbit and the envelope of the primary. Furthermore, we cannot predict the relationship between pre-CE and post-CE populations \citep{Izzard2012,DeMarco2017}. The types of advances that can be expected from exploration of the  APOGEE-GALEX-{\it Gaia} catalog are demonstrated by our analysis of 45 previously classified WDMS binaries (identified by SDSS and LAMOST) that have also been observed by APOGEE \citep{Corcoran2020}. Based on that pilot survey, \cite{Corcoran2020} show that the wide binary stars in the sample have an MDF that is significantly skewed to lower metallicities than the PCEB stars. That sample was mainly limited to WD - M-dwarf pairs.  Here we extended the \cite{Corcoran2020} study to the much larger APOGEE-GALEX-\emph{Gaia} catalog to characterize many more previously {\it unknown} WDMS systems, as well as systems with evolved secondaries.

A particularly contribution here is the sub-catalog containing 252 systems with well-constrained orbits from The Joker \citep{Price2017} (see \S\ref{time}). For a larger sample than \cite{Corcoran2020} and across the Hertzsprung-Russell diagram, we find that the a sample dominated by WB ($P > 100$ days) is skewed to lower metallicities than a sample dominated by PCEB stars ($P < 100$ days). Moreover, a detailed comparison between a sample dominated by WD binaries and other samples containing mainly single stars (see \S\ref{Proper_HR}) shows that there is a prominent difference between the MDFs for the most luminous giants, $M_{H} < -3.0$ (see Fig.~\ref{CMD_WD_binaries_Lumi}). This range of absolute magnitude might contained a large fraction of RG and AGB stars, where the stellar radii can reach sizes of several astronomical units. Binary interactions can have an impact on the intrinsic properties of an evolved star: It can alter the pulsations, the mass-loss efficiency and geometry, the dust formation processes, and the circumstellar envelope morphology. Binary interaction can even play a dominant role in determining the ultimate fate of the object, i.e., the formation of CVs, SNe Ia, Barium stars, gravitational wave sources, etc. \citep{J&B2017,Oomen2018}. Our findings support the scenario where there is an enrichment of a system’s surface chemistry during the CE phase. While in this pilot study we only explored the overall metallicity measured from the APOGEE spectra, in future work we will study in detail the abundance differences in the WD binary identify sample created here for other, individual chemical elements (the abundances of more than 15 elements exist in the APOGEE catalog, including C, N, O, Na, Mg, Al, Si, S, K, Ca, Ti, V, Mn, Fe, and Ni). 


\section{Summary}
\label{summary}

In this paper we present a systematic search for compact binary star systems containing white dwarfs, created by harnessing information contained in the spectroscopic catalog of the APOGEE project \citep{2017AJ....154...94M}, matched with the data from the {\it Gaia} \citep{Lindegren2018}  and GALEX \citep{Bianchi2017} space missions. The results of our investigation of these systems are as follows: 
\begin{enumerate}
\item We have created (\S2) the APOGEE-GALEX-Gaia Catalog (AGGC). Systems with $T_{\rm eff}$ $<$ 6000 K and $(FUV - NUV)_{\circ}< 5$ are the most compelling for our purposes. Based on these selections, we have identified 3,414 APOGEE sources that are potential WD binary candidates having F-M spectral type companions. 
\item We use the empirical spectral energy distribution from the UV to the IR for the AGGC sample to derive system parameters like the WD temperature and radius (\S3). The radii diagram (\S4, Fig.~\ref{WD_radii_MS}) provides a useful criterion for creating a relatively clean sample of WD binary candidates within the AGGC sample, where a total of 1,806 stars have WD radii that fall below 25 R$_{\Earth}$ and constitute a more reliable sample of WD binary candidates. 
%
\item The most luminous ($M_H < -3.0$) secondaries in our WD binary candidate sample clearly show a different MDF with respect to a control sample dominated by single stars of similar luminosity (\S5.1, Fig.~\ref{CMD_WD_binaries_Lumi}). This may have to do with the effects of binary companions on the chemical evolution of the AGB population --- e.g., in how many are converted into carbon stars. We have previously shown how symbiotic stars identified in our catalog show enhancement in carbon abundance \citep[e.g.,][]{Lewis2020,Washington2021}.
\item We also highlight (\S5.1) the existence of a sub-subgiant population \citep[e.g.,][]{Beloni1998,Mathieu2003} in Figure~\ref{CMD_WDMS} and Figure~\ref{CDF_temp_meta}. These objects are X-ray sources and photometric variables. Where binary status is known, they are often found to be close binary systems with orbital periods on the order of 1-10 days \citep{Leiner2017,Geller2017}. In this exercise we found that the orbital period associated to these objects ranges from a few days to $\sim$ 20 days (see Fig.~\ref{CDF_temp_meta}), in agreement with the literature. The sub-subgiants in this sample potentially contain a WD companion, this sample could be an opportunity to make substantial new progress in understanding these intriguing systems with additional detailed modeling. 
\item The $\Delta$RV$_{max}$ cumulative fraction for the WD binaries is skewed towards larger $\Delta$RV$_{max}$ values with respect to the APOGEE DR17 control sample (\S5.2, Fig.~\ref{WDB_cum}). This result suggest shorter periods for these systems, and an evidence for drastic loss of angular momentum associated with the formation of the WD. We also investigated 252 potential WD binaries with estimated orbital periods. From these we find a large fraction to show an orbital period of a few days, typical of binary systems whose orbits have
circularized (Fig.~\ref{WDB_P_CDF}). From the CDF of two different samples selected using the orbital period, we found (Fig.~\ref{Fe_period}) that the sample dominated by wide binaries ($P > 100$ days) is skewed to lower metallicities than the sample dominated by PCEB stars ($P < 100$ days). This finding suggest an enrichment of a systems surface chemistry during the CE phase. 
%
\end{enumerate}

The AGGC is a rich resource for investigating the evolution of WD binaries across the H-R diagram.  Here we have only touched various avenues that are ripe for further development. Among the additional available tools that we intend to exploit in our future efforts are the 
more than 15 elements derived in the APOGEE catalog for the WD binary sample, and looking more deeply into the orbital properties of the systems, beyond simple periods.


\acknowledgments
We thank the referee for comments that helped improve this paper.
BA acknowledges support from a Chr\'etien International Research Grant from the American Astronomical Society. BA and SRM acknowledge support from National Science Foundation grant AST-1616636. 
The authors acknowledge helpful discussions with E.\ Leiner regarding sub-subgiants and for sharing SSG evolution models.
Funding for the Sloan Digital Sky Survey IV has been provided by the Alfred P. Sloan Foundation, the U.S. Department of Energy Office of Science, and the Participating Institutions. SDSS-IV acknowledges support and resources from the Center for High-Performance Computing at the University of Utah. The SDSS web site is www.sdss.org. SDSS is managed by the Astrophysical Research Consortium for the Participating Institutions of the SDSS Collaboration including the Brazilian Participation Group, the Carnegie Institution for Science, Carnegie Mellon University, the Chilean Participation Group, the French Participation Group, Harvard-Smithsonian Center for Astrophysics, Instituto de Astrof\'isica de Canarias, The Johns Hopkins University, Kavli Institute for the Physics and Mathematics of the Universe (IPMU) / University of Tokyo, the Korean Participation Group, Lawrence Berkeley National Laboratory, Leibniz Institut f\"ur Astrophysik Potsdam (AIP), Max-Planck-Institut f\"ur Astronomie (MPIA Heidelberg), Max-Planck-Institut f\"ur Astrophysik (MPA Garching), Max-Planck-Institut f\"ur Extraterrestrische Physik (MPE), National Astronomical Observatories of China, New Mexico State University, New York University, University of Notre Dame, Observat\'orio Nacional / MCTI, The Ohio State University, Pennsylvania State University, Shanghai Astronomical Observatory, United Kingdom Participation Group, Universidad Nacional Aut\'onoma de M\'exico, University of Arizona, University of Colorado Boulder, University of Oxford, University of Portsmouth, University of Utah, University of Virginia, University of Washington, University of Wisconsin, Vanderbilt University, and Yale University. This research has made use of the SIMBAD database,
operated at CDS, Strasbourg, France. This publication makes use of VOSA, developed under the Spanish Virtual Observatory project supported by the Spanish MINECO through grant AyA2017-84089.
VOSA has been partially updated by using funding from the European Union's Horizon 2020 Research and Innovation Programme, under Grant Agreement nº 776403 (EXOPLANETS-A). This research has made use of NASA’s Astrophysics Data System Bibliographic Services (NASA ADS). This research made use of Astropy,\footnote{http://www.astropy.org} a community-developed core Python package for Astronomy.




\end{document}